\documentclass[lineno]{jfm}

\usepackage{graphicx}
\usepackage{newtxtext}
\usepackage{newtxmath}
\usepackage{natbib}
\usepackage[export]{adjustbox}
\usepackage{caption} 
\usepackage{subcaption} 
\usepackage{wasysym} 
\usepackage{multirow} 
\usepackage{hyperref}
\usepackage[justification=raggedright,singlelinecheck=false]{caption}

\hypersetup{
    colorlinks = true,
    urlcolor   = blue,
    citecolor  = black,
}

\newcommand{\RomanNumeralCaps}[1]
\linenumbers

\title{Coherent-structure dynamics in wall turbulence from state-space trajectories}

\author{Emma Lenz\aff{1}
  \corresp{\email{elenz@caltech.edu}},
  Ahmed Elnahhas\aff{2}
 \and H. Jane Bae\aff{1}}

\affiliation{\aff{1}Lynn Booth and Kent Kresa Department of Aerospace, California Institute of Technology, Pasadena, CA 91125, USA
\aff{2}Center for Turbulence Research, Stanford University, Stanford, CA 94305, USA}

\begin{document}

\maketitle

\begin{abstract}
    We identify dynamical processes in the near-wall region of turbulent wall-bounded flow by representing coherent-structure energy as trajectories in a low-dimensional state space and analyzing recurring trajectory patterns using network motifs. Direct numerical simulations (DNS) are performed for three configurations: a minimal flow unit (MFU) at $Re_\tau\approx180$ to isolate the self-sustaining process (SSP), a full-scale channel at the same Reynolds number to study interactions between near-wall structures, and an MFU at $Re_\tau\approx2200$ to investigate near-wall/outer-layer coupling. Proper orthogonal decomposition (POD) is used to identify modes corresponding to streaks, rolls, and meandering structures of the SSP, and the flow is  projected onto these modes to track their energy over time. Motif analysis then identifies statistically significant dynamical pathways in the resulting state-space trajectories. Several motifs are common to all three configurations, indicating robust near-wall dynamics across different background flows. While motifs associated with the classical SSP are recovered, equally prominent motifs with no direct SSP analogue are also identified, demonstrating that preferred near-wall dynamics extend beyond the canonical regeneration cycle. Conditioning the $Re_\tau\approx2200$ MFU on outer-layer energy further shows that quiescent outer-layer states produce dynamics resembling those of the low-Reynolds-number full channel, whereas energetic outer-layer states promote high-energy bursting events. These results demonstrate that near-wall turbulence evolves along preferred dynamical pathways whose prevalence is modified, but not eliminated, by outer-layer activity.
\end{abstract}


\section{Introduction}
\label{sec:introduction}

Turbulent wall-bounded flows are characterized by a wide range of interacting length scales, yet viewing turbulence as a system of coherent structures reveals organized patterns amidst the apparent chaos. Coherent structures are typically defined as regions of high energy or stress that persist over a range of spatial and temporal scales, and their interactions drive the key dynamical processes of wall-bounded turbulence \citep{kline1967structure, cantwell1981organized, robinson1991coherent, kawahara2009theoretical}. Despite extensive characterization of these structures through analysis, simulation, and experiment, a central challenge remains: identifying the recurring dynamical processes that govern turbulence in the presence of interactions across multiple spatial and temporal scales. In this work, we address this challenge by identifying recurring dynamical processes through the evolution of energy contained in coherent structures, with particular emphasis on how these processes relate to established mechanisms in near-wall turbulence.

\begin{figure}
    \centering
    \includegraphics[width=\linewidth,trim={5cm 7cm 3cm 2cm},clip]{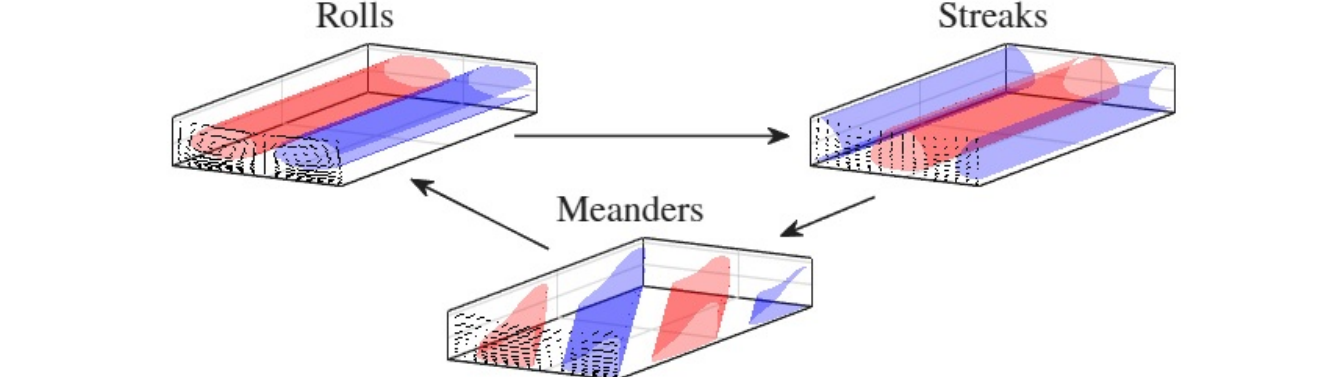}
    \caption{Schematic of the self-sustaining process. Roll mode shows local maxima in the cross-stream velocity, streak mode shows isocontours of streamwise velocity, and meander mode shows isocontours of vorticity.}
    \label{fig:sspcartoon}
\end{figure}

A foundational dynamical mechanism in near-wall turbulence is the SSP, which describes a quasi-cyclic interaction between streamwise rolls, streaks, and their instabilities \citep{panton2001overview}. A schematic of this process is shown in figure \ref{fig:sspcartoon}. Streamwise rolls redistribute momentum and generate streaks, whose subsequent instability produces oblique structures (meanders) that regenerate the rolls \citep{waleffe1995hydrodynamic, waleffe1997self}. The SSP is conceptually appealing because it provides a closed dynamical loop directly connected to the governing equations, offering a mechanistic explanation for how near-wall turbulence is sustained.

The SSP can be observed most clearly in simplified flow configurations that isolate the essential near-wall dynamics. MFUs — channels truncated to the smallest streamwise and spanwise dimensions that sustain healthy near-wall turbulence \citep{jimenez1991minimal, flores2010hierarchy} — remove confounding large-scale interactions and allow the SSP cycle to emerge in a relatively clean and persistent form \citep{hamilton1995regeneration, jimenez1999autonomous, jimenez2004large}. The spanwise dimension of an MFU corresponds to the characteristic streak spacing of approximately 100 viscous units, a scale that plays a critical role in sustaining turbulence and has been central to studies of streak generation \citep{waleffe1993origin, butler1992three, jang1986origin}. In the streamwise direction, the finite domain causes structures with very long streamwise wavelengths to manifest as streamwise-constant modes, which contribute significantly to near-wall dynamics \citep{aubry1988dynamics, hamilton1995regeneration, jimenez1999autonomous}.
The importance of these scales is underscored by the fact that their removal leads to relaminarization, indicating that they form the minimal set of structures required to sustain turbulence.

Equation-based analyses and numerical experiments have further illuminated the energetic mechanisms underlying the SSP, including linear streak amplification \citep{kim2000linear, del2006linear, lozano2021cause} and nonlinear streak breakdown and roll regeneration \citep{schoppa2002coherent, bae2021nonlinear, ballouz2025transient}. These insights have enabled the construction of low-dimensional models that reproduce the SSP cycle \citep{waleffe1997self, dauchot2000phase, moehlis2004low}, as well as the discovery of exact coherent structures  \citep{waleffe1998three, waleffe2001exact, moehlis2005periodic, park2018bursting, jimenez1987coherent, kawahara2001periodic}, which clarified the SSP's energetic mechanisms and their connections to bursting \citep{jimenez2005characterization, park2018bursting} and relaminarization \citep{kawahara2006unstable}. 

However, the clarity of the SSP picture diminishes as one moves from these idealized systems to full turbulent flows. In realistic configurations, even at similar Reynolds numbers as the MFU, the coexistence of a wide range of interacting length and time scales in the near-wall region (intra-layer interactions) obscures the clean quasi-cyclic behavior observed in the MFU. 
As a result, studies of full turbulence have relied on alternative definitions of coherent structures and indirect methods to infer dynamical processes. Coherent structures can be extracted from large flow fields by applying a threshold to a quantity of interest, such as the discriminant criterion \citep{blackburn1996topology, del2006self} or Reynolds stress \citep{lozano2012three}, and tracked over time to reveal patterns in their growth, decay, and interactions \citep{lozano2014time, bae2021life}. Coherent structures also form the basis of reduced-order models, where structures are defined via modal decompositions such as POD and dynamic mode decomposition; with an appropriate choice of modes, these models can reproduce statistics of full-scale turbulence \citep{aubry1988dynamics, jimenez1994structure, holmes1997low, rowley2017model}. 
While these approaches characterize coherent structures effectively, they do not directly reveal the recurring dynamical processes that connect them.

At higher Reynolds numbers, this challenge becomes more pronounced due to the emergence of outer-layer dynamics and a hierarchy of interacting structures (inter-layer interactions). While near-wall turbulence can be sustained independently of the outer flow \citep{jimenez1999autonomous}, the logarithmic and outer regions exhibit their own forms of self-sustaining dynamics distributed across a range of scales \citep{del2006self, hwang2010self, hwang2011self, hwang2016self, cossu2017self, bae2021nonlinear}. MFUs at higher Reynolds numbers can isolate individual active scales \citep{flores2010hierarchy} or a small number of discrete scales \citep{doohan2021minimal}, providing evidence that SSP-like dynamics extend beyond the near-wall region. In fully developed turbulence, however, these processes are embedded within a hierarchy of interacting structures, where energy is exchanged across scales and wall-normal locations \citep{cho2018scale, lee2019spectral}. Consequently, identifying the dynamical processes governing the near-wall region becomes increasingly challenging.

The influence of the outer layer on near-wall turbulence has been studied extensively in the context of superposition and amplitude modulation. Townsend's attached eddy hypothesis proposes that wall turbulence consists of a hierarchy of self-similar wall-attached eddies whose characteristic size scales with the distance from the wall, such that near-wall turbulence may be viewed as the superposition of near-wall structures and larger attached eddies \citep{townsend1976structure, marusic2019attached}. However, \cite{morrison2007interaction} showed that linear superposition alone cannot explain inner–outer interactions. Subsequent studies demonstrated that large-scale outer structures also modulate the amplitude of near-wall fluctuations \citep{hutchins2007evidence, hutchins2007large,mathis2009large}, leading to predictive models of near-wall turbulence based on outer-layer information \citep{marusic2010predictive}. 
These results indicate that outer-layer motions influence not only the amplitude of near-wall fluctuations, but also the dynamical evolution of near-wall coherent structures. 

An important question is whether the dynamical processes governing near-wall turbulence exhibit universality when expressed in terms of characteristic near-wall length and time scales.
Mean velocity profiles and energy spectra are known to collapse near the wall across Reynolds numbers \citep{millikan2938critical, wosnik2000theory, jimenez2013near}, although deviations emerge at high Reynolds numbers, including a secondary peak in the streamwise fluctuation intensity whose wall-normal location scales as $Re^{1/2}$ \citep{morrison2004scaling, smits2011high, lee2015direct}. Likewise, coherent structures in the outer layer show self-similar behavior, but this does not extend to near-wall structures \citep{townsend1976structure, del2006self, lee1990structure}. Whether the underlying dynamical processes, rather than statistical quantities alone, exhibit similar universality remains largely unexplored. Addressing this question requires a framework that identifies recurring dynamical processes in a manner that permits direct comparison across different flow configurations.

Taken together, these observations highlight a key challenge in wall turbulence. While the SSP provides a compelling conceptual framework for near-wall turbulence, its direct observability becomes increasingly limited in realistic flows where coherent structures interact across a broad range of spatial and temporal scales. This motivates the need for approaches that identify dynamical processes without requiring canonical SSP cycles to be directly observable. 

In this work, we represent the energy contained in coherent structures as trajectories through a low-dimensional state space and apply graph-theoretic motif analysis to identify statistically significant trajectory motifs, in which overrepresented trajectory patterns are identified relative to randomized sequences with the same lower-order  statistics \citep{iacobello2021review, elnahhas2024dynamics}. This motif analysis identifies statistically overrepresented trajectory patterns that are consistent with recurring dynamical processes that cannot be inferred from averaged trajectories or individual state transitions alone, enabling systematic comparisons across different flow configurations. Applying the framework to minimal and full turbulent channels spanning low and high Reynolds numbers allows us to distinguish dynamical processes associated with the classical SSP from additional recurring processes and to quantify how intra-layer and outer-layer interactions modify near-wall dynamics. Existing turbulence analyses identify coherent structures or average cycles; we instead identify statistically preferred dynamical pathways using trajectory motif analysis.
The paper is organized as follows: the methodology is described in \S\ref{sec:methods}, including the numerical datasets (\S\ref{sec:numericaldatasets}), coherent structure identification (\S\ref{sec:POD}), state space framework (\S\ref{sec:statespace}), and motif significance metric (\S\ref{sec:motif}); \S\ref{sec:results} presents trajectory analysis and motif results, including high- and low-energy outer-layer conditioning for the $Re_\tau \approx 2200$ MFU; \S\ref{sec:discussion} relates the results to well-known dynamical processes; \S\ref{sec:conclusion} summarizes key findings.


\section{Methods}
\label{sec:methods}

\begin{figure}
    \centering
    \includegraphics[width=\textwidth,trim={0cm 12cm 0cm 0cm},clip]{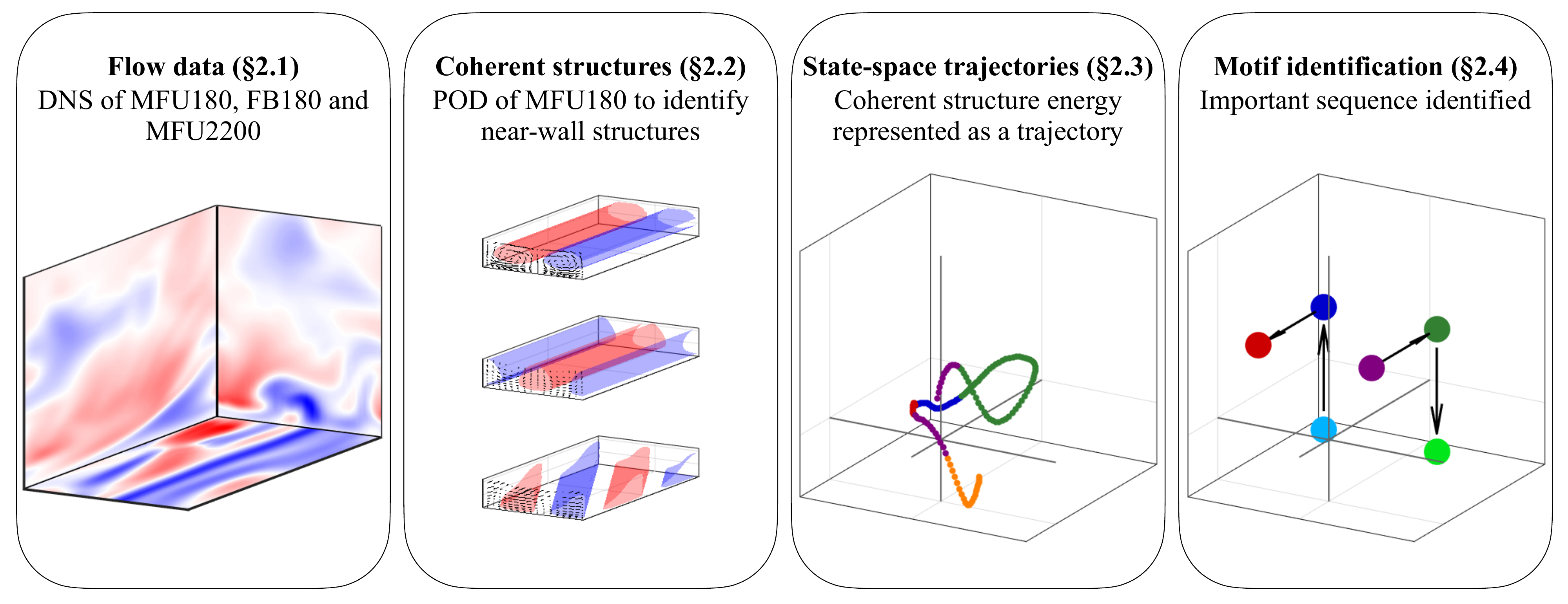}
        
    \caption{Summary of methods. In \S\ref{sec:numericaldatasets}, DNS is performed for the MFU180, FB180, and MFU2200, and sub-domains of consistent sizes are tracked. In \S\ref{sec:POD}, POD is performed on the near-wall region of the MFU180, and coherent structures reflecting the SSP are identified and their energy is tracked. In \S\ref{sec:statespace}, energy in the three identified modes is represented as a trajectory through state space, and the series of octants traced by the trajectory creates a sequence. In \S\ref{sec:motif}, motif analysis is used to quantify the significance of sub-sequences of octants, identifying important dynamical pathways.} 
    \label{fig:overviewfigure}
\end{figure}

The methodology consists of four phases. In the first phase (\S\ref{sec:numericaldatasets}), DNS of turbulent channel flow of different dimensions are performed, and advecting sub-domains are used to extract near-wall data consistently in the channels. In the second phase (\S\ref{sec:POD}), POD is used to identify energetic coherent structures, and the energy contained in the structures is tracked over time. In the third phase (\S\ref{sec:statespace}), the energy in coherent structures is represented as a trajectory through a state space. Finally, in the fourth phase (\S\ref{sec:motif}), a graph-theoretic motif analysis is applied to extract and quantify significant dynamical patterns from the trajectory. A preliminary version of this methodology can be found in \cite{elnahhas2024dynamics}, though here we extend the approach to study the effect of outer-layer structures and inter-layer interactions. A visual summary of the methods is shown in figure \ref{fig:overviewfigure}.

Throughout this work, we use $x$, $y$, and $z$ to denote the streamwise, wall-normal, and spanwise directions, respectively. The corresponding velocity fluctuations are given by $u$, $v$, and $w$, and $\mathbf{u} = [u,v,w]^*$, where $*$ denotes conjugate transpose. The mean streamwise velocity $U(y)$ is averaged over the homogeneous directions ($x$ and $z$) and time $t$, and mean wall-normal and spanwise velocities are taken to be zero. The friction Reynolds number is defined as $Re_\tau=\delta u_\tau/\nu$, where $\delta$ is the channel half-height, $u_\tau$ is the friction velocity, and $\nu$ is the kinematic viscosity. The superscript $+$ denotes quantities scaled in viscous units, i.e. by $u_\tau$ and $\nu$.

\subsection{Numerical experiments}
\label{sec:numericaldatasets}
\subsubsection{Direct numerical simulations of turbulent channel flow}


\begin{figure}
    \centering
    \includegraphics[width=\textwidth,trim={3cm 2cm 3cm 4.5cm},clip]{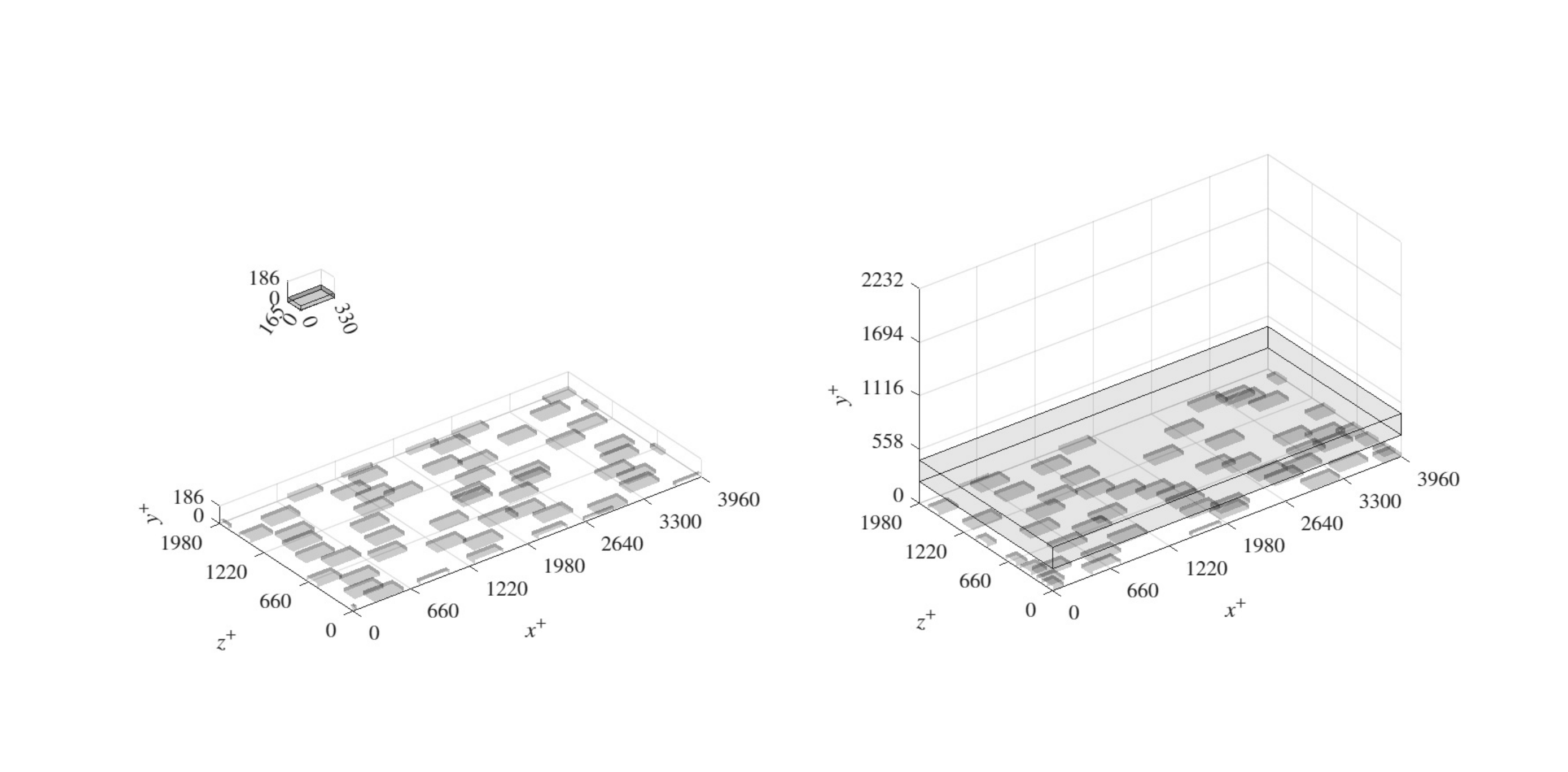}
        \begin{flushleft}
        \vspace{-4cm}
        \hspace{.6cm}
        (a)
        \vspace{4cm}
        \\
        \vspace{-1.3cm}
        \hspace{0.3cm}
        (b)
        \hspace{6.2cm}
        (c)
        \vspace{1.3cm}
    \end{flushleft}
    \caption{Diagrams of the DNS datasets: (a) MFU180, (b) FB180, and (c) MFU2200. Small shaded boxes show the sub-domains over which energy in $r$, $s$, and $m$ are calculated. For the FB180 and MFU2200, the sub-domains advect over time, and an instantaneous snapshot of their locations is shown. The large shaded box in the MFU2200 shows the outer layer, over which outer layer turbulent kinetic energy $K_{OL}$ is computed.}
    \label{fig:channeldigrams}
\end{figure}

We consider three configurations of turbulent channel flow, each designed to isolate different types of dynamical interactions (see figure \ref{fig:channeldigrams}). We refer to interactions between near-wall coherent structures as intra-layer interactions, and interactions between near-wall coherent structures and outer-layer dynamics as inter-layer interactions.

The first configuration is an MFU at Reynolds number $Re_\tau\approx180$ (MFU180), with streamwise and spanwise dimensions $L_x/\delta \approx \pi/2$ and $L_z/\delta \approx\pi/4$. This is the smallest possible domain that can sustain turbulence, containing at most one low-speed and high-speed streak pair or one pair of counter-rotating rolls \citep{jimenez1991minimal}. By restricting the length scales present, the MFU180 isolates the near-wall coherent structures from both intra-layer and inter-layer interactions, allowing its dynamical patterns to be observed in isolation.  

The second configuration is a full-scale channel at $Re_\tau\approx180$ (FB180), with dimensions $L_x\approx6\pi\delta$ and $L_z\approx3\pi\delta$. The large domain accommodates many coherent structures that interact with one another and with larger scales that are absent in the MFU180, making this configuration well-suited for studying intra-layer interactions in the near-wall region. However, since the Reynolds number is low, no distinct outer-layer structures are present and inter-layer interactions are absent.

The third configuration is an MFU at $Re_\tau\approx2200$ (MFU2200), with the same outer-scaled dimensions as the MFU180 ($L_x/\delta \approx \pi/2$ and $L_z/\delta \approx \pi/4$)\citep{flores2010hierarchy} and the same viscous-scale dimensions as the FB180 ($L_x^+=3960$ and $L_z^+=1980$). This domain size is chosen to contain the same range of near-wall length scales present in the FB180, while introducing the effect of an outer-layer dynamic, making this configuration well-suited for studying both intra-layer and inter-layer interactions.

DNS of the incompressible Navier--Stokes equations are performed using a staggered second-order central finite difference scheme with a fractional step method. Time advancement is performed using a third-order Runge--Kutta scheme. A periodic boundary condition is applied in the streamwise and spanwise directions, and the no-slip and no-penetration boundary condition is applied at the walls. The grid is uniform in the streamwise and spanwise directions, with $\Delta x^+=\Delta z^+=5.16$. In the wall-normal direction, the grid is stretched by a hyperbolic tangent function, with a minimum spacing of $\Delta y^+_{min}=0.17$ and a maximum spacing of $\Delta y^+_{max}=7.5$ in the MFU180 and FB180, and $\Delta y^+_{max}=11.7$ in the MFU2200. In the MFU2200, wall-normal grid spacing below $y^+\approx180$ matches that of the MFU180 and FB180 so that flow fields can be directly compared without interpolation. 
This solver has been validated and used in several previous studies \citep{bae2018turbulence, bae2019dynamic, bae2021nonlinear, bae2021life, ballouz2025transient}. 

\begin{figure}
     \centering
     \begin{subfigure}[b]{0.49\textwidth}
         \centering
         \caption{}
         \includegraphics[width=\textwidth,trim={0cm .5cm 2cm 0.5cm},clip]{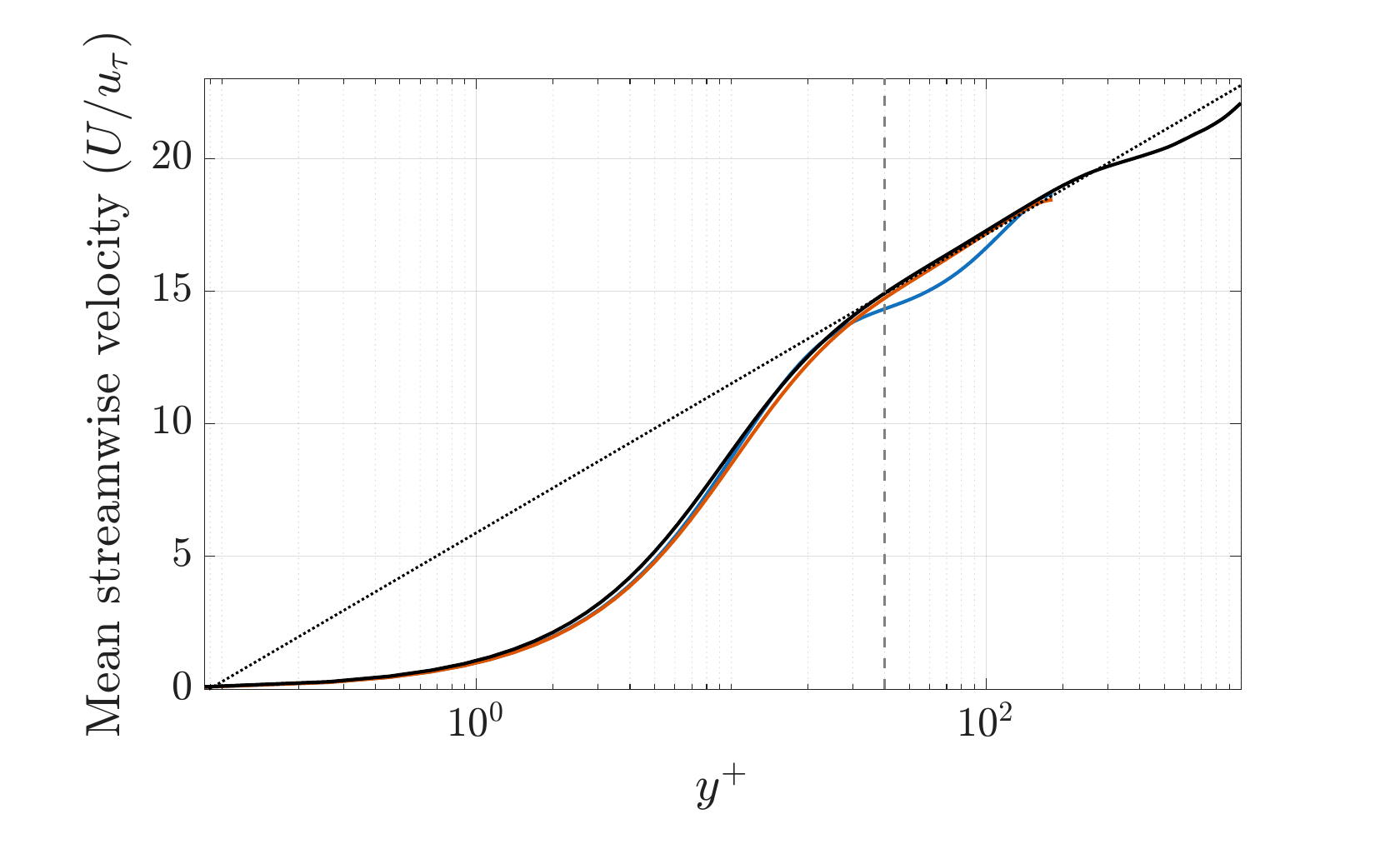}
         \label{fig:universality_mean}
     \end{subfigure}
     \begin{subfigure}[b]{0.49\textwidth}
         \centering
         \caption{}
         \includegraphics[width=\textwidth,trim={0cm .5cm 2cm 0.5cm},clip]{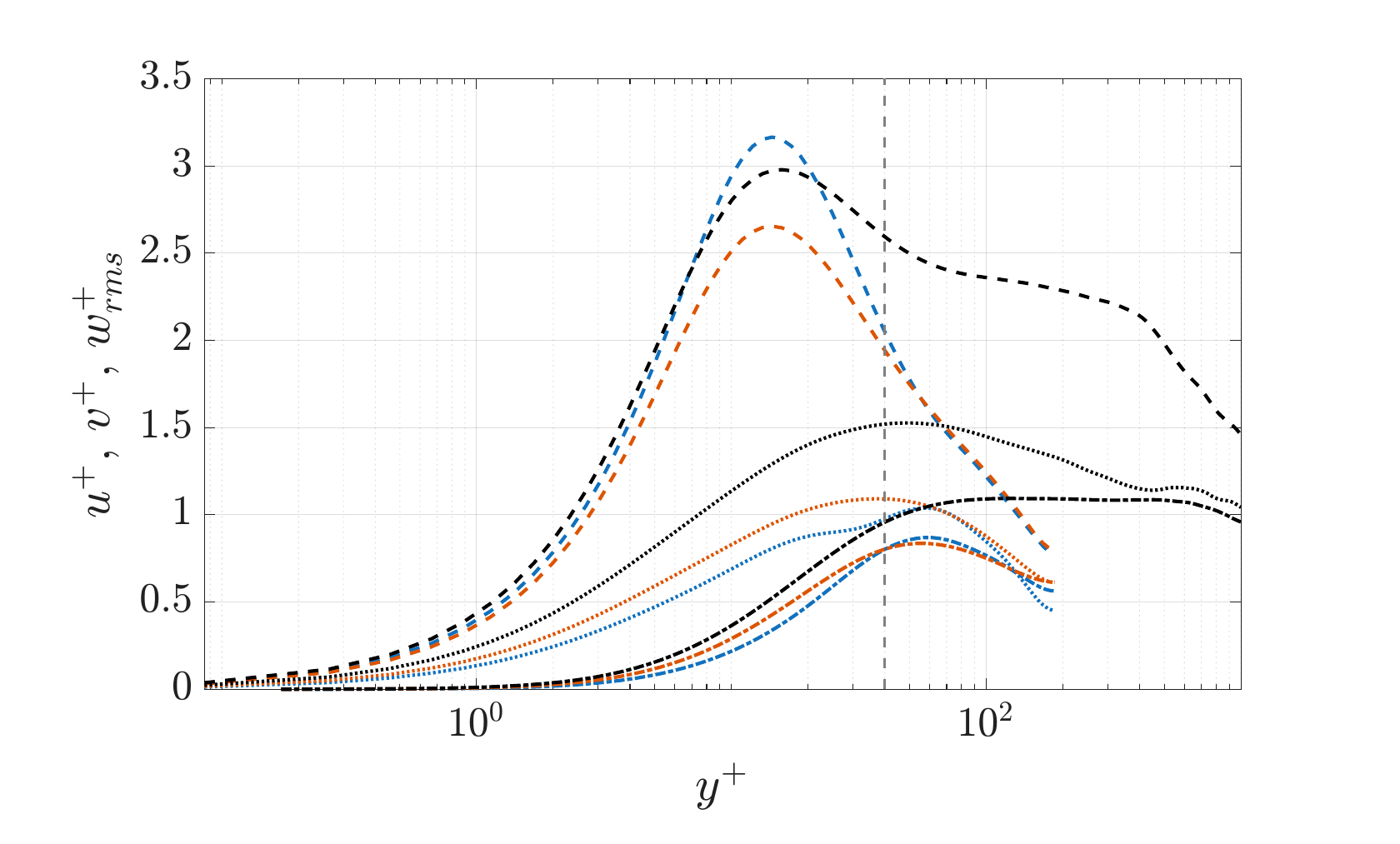}
         \label{fig:universality_rms}
     \end{subfigure}
        \caption{(a) Mean streamwise velocities for the MFU180 (blue), FB180 (red), and MFU2200 (black). Grey dashed line shows the boundary of the near-wall region, defined as $y^+\le40$. Dotted line is the log law, $U^+=\frac{1}{\kappa}y^++B$, with $\kappa=0.41$ and $B=5$. (b) Turbulence intensities of the streamwise (dashed), wall-normal (dot-dashed), and spanwise (dotted) components of the same datasets.}
        \label{fig:universality}
\end{figure}

The mean streamwise velocity profiles and velocity fluctuation intensities from the three simulations are shown in figure \ref{fig:universality}. We observe that the mean velocity profiles are very similar near the wall for all three cases, and while the fluctuation intensities have different magnitudes, the curves have similar features and peak at the same wall-normal location in viscous units.

\subsubsection{Sub-domain tracking}


To consistently observe the dynamical evolution of coherent structures in all three configurations, we limit our observation domain to the size of the MFU180 in the streamwise and spanwise directions. In addition, the domain is limited to $0<y^+<40$ in the wall-normal direction to study the near-wall region, resulting in sub-domains of size $(L^+_{x,SD}, L^+_{y,SD}, L^+_{z,SD}) = (330, 40, 165)$.


In the FB180 and MFU2200, near-wall structures are advected by larger-scale motions, so the sub-domains are translated in the streamwise and spanwise directions at each timestep to follow these structures. The translation is chosen to maximize the correlation of the flow field between consecutive timesteps, and the translation displacements for each sub-domain are given by


\begin{equation}
    (\Delta x_{SD},\Delta z_{SD}) = \underset{\Delta x, \Delta z}{\operatorname{argmax}}\left\{ \underset{SD}{\operatorname{corr}} \big{(} u(x+\Delta x,y^+=40,z+\Delta z,t+\Delta t), \ u(x,y^+=40,z,t)\big{)} \right\},
\end{equation}
where $\underset{SD}{\operatorname{corr}}\big{(}\cdot,\cdot\big{)}$ is the correlation function over a sub-domain. The correlation is computed at $y^+=40$ — a wall-normal location at which energetic coherent structures are known to be minimally affected by shearing \citep{lozano2014time}. Consequently, the sub-domains advect with large-scale coherent structures. To confirm that the sub-domains capture dynamical cycles of turbulent energy, we compute and analyze the turbulent production and dissipation, which are discussed further in appendix~\ref{appA}.


192 evenly-spaced sub-domains are initialized in the FB180 and MFU2200 to improve statistical sampling. As the sub-domains advect, however, they tend to converge toward the same large-scale structures, eventually overlapping and advecting at the same velocity. To prevent this, sub-domains are reset to their initial positions after $\Delta t^+=372$, corresponding to two eddy turnover times. This yields continuous temporal trajectories of duration $\Delta T_{traj}^+=372$, which form the basis of the state space analysis described in the following sections.

\subsubsection{Outer-layer energy}

To study the effect of the outer-layer dynamics of the MFU2200 on the near-wall region, we employ conditioning on the outer-layer energy. We define the outer region as $200 \nu/u_\tau < y < 0.2 \delta$ in the wall-normal direction, overlapping with the nominal log layer \cite{marusic2013logarithmic}, and spanning the full streamwise and spanwise extent of the MFU2200. 
This region has dimensions $(L_{x,OL}^+, L_{y,OL}^+, L_{z,OL}^+) = (3960, 246, 1980)$ and contains large-scale energetic coherent structures \citep{kim1999very, agostini2014influence}.


Outer-layer energy is then defined as the time-averaged turbulent kinetic energy over each trajectory in the outer region,
\begin{equation}
K_{OL} = \frac{1}{2\Delta T}\int_{0}^{\Delta T}\int_0^{L_{z,OL}}\int_{200\frac{\nu}{u_\tau}}^{0.2\delta}\int_0^{L_{x,OL}} \left( u^2 + v^2 + w^2\right)\,\mathrm{d}x\,\mathrm{d}y\,\mathrm{d}z\,\mathrm{d}t,
\end{equation}
where the spatial integral is taken over the outer region and the temporal integral is taken over the trajectory duration $\Delta T^+ = 372$. 
Since the outer-layer energy evolves slowly relative to the near-wall region, averaging over the full trajectory duration ensures that the classification is insensitive to short-time fluctuations within a single trajectory while capturing the overall energy affecting the near-wall dynamics. By conditioning on outer-layer energy we observe the effects of footprinting and amplitude modulation on the near-wall dynamics, though since we do not filter or subtract the footprints of large outer-layer structures, the results do not isolate the effects of either footprinting or amplitude modulation and instead reflect their combined effects.

\subsection{Tracking energy in SSP structures}
\label{sec:POD}


Next, we identify energetic coherent structures that play a role in the SSP and track their energy over time. We perform a POD of the near-wall region using data from the MFU180 \citep{lumley1981coherent}, with domain $(L_{x,SD}^+,L_{y,SD}^+,L_{z,SD}^+)=(330,40,165)$. POD identifies orthogonal modes that optimally capture the variance of a dataset; here, we seek modes that optimally capture kinetic energy. To this end, we define the inner product between two vector quantities $\boldsymbol{\xi}(x,y,z)$ and $\boldsymbol{\zeta}(x,y,z)$ as
\begin{equation}
\left\langle \boldsymbol{\xi} , \boldsymbol{\zeta} \right\rangle = \frac{1}{L_{x,SD} L_{y,SD} L_{z,SD}}\int_{0}^{L_{z,SD}}\int_0^{L_{y,SD}}\int_{0}^{L_{x,SD}} \boldsymbol{\zeta}^*(x,y,z) \cdot \boldsymbol{\xi}(x,y,z)\,\mathrm{d} x\, \mathrm{d} y\, \mathrm{d} z,
\end{equation}
which corresponds to the kinetic energy norm, i.e., 
\begin{equation}
    \frac{1}{2}\|\mathbf{u}\|^2 = \frac{1}{2}\left\langle \mathbf{u}, \mathbf{u} \right\rangle = K,
\end{equation}
where $K$ is the instantaneous kinetic energy.

Since the channel is homogeneous and periodic in the streamwise and spanwise directions, the POD modes in those directions are Fourier modes \citep{holmes2012turbulence}. We therefore Fourier transform the flow field in $x$ and $z$ such that $\widehat{\mathbf{u}}(k_x,k_z;y,t)$ denotes Fourier-transformed velocity field $\mathbf{u}$, where $k_x$, $k_z$ are wavenumbers reflecting the number of periods in each direction. The POD modes are then obtained by taking the singular value decomposition in the wall-normal direction,
\begin{equation}
\widehat{\mathbf{u}}(k_x,k_z;y,t) = \sum_{n=1}^{N} \boldsymbol{\phi}(k_x,k_z,n;y) \ \sigma(k_x,k_z,n)\ \boldsymbol{\psi}^*(k_x,k_z,n;t),
\end{equation}
where $\boldsymbol{\phi}$ are the left singular vectors (spatial modes), $\boldsymbol{\psi}$ are the right singular vectors (temporal coefficients), $\sigma$ are the singular values sorted in descending order, and $n$ indexes the modes. The mode $n=1$ thus corresponds to the most energetic spatial structure for each wavenumber pair $(k_x, k_z)$. 

The simulation resolution captures $N_x=64$ modes in the streamwise direction, $N_z=32$ modes in the spanwise direction, and $N=3N_y=120$ modes in the wall-normal direction, resulting in a total of $N_{total}=N_x N_y N=2.5\times10^5$ modes. We denote all singular vectors for all values of $k_x = 0,...,N_x-1$, $k_z = 0,..., N_z-1$, and $n=1,2, ..., N$ with a single index based on $\sigma$, such that the set of all modes can be expressed as $\{\phi(k_{x,q},k_{z,q},n_q;y) | q = 1,2,...,N_{total} \}$ and $\sigma_1 > \sigma_2 > \ ... \ > \sigma_{N_{total}} \geq 0 $. For brevity, a mode with wavenumbers $(k_{x,q}, k_{z,q})$ and index $n_q$ is referred to as mode $q$, with $\boldsymbol{\phi}(k_{x,q},k_{z,q},n_q;y)$ abbreviated as $\boldsymbol{\phi}_q(y)$. The energy contained in mode $q$ at time $t$ is
\begin{equation}
\label{eq:modeenergy}
E_q(t) = \frac{1}{2}|\beta_q(t)|^2,
\end{equation}
where $\beta_q(t)$ is the complex projection coefficient obtained by projecting the flow field onto mode $q$,
\begin{equation}
\label{eq:beta}
\beta_q(t) = \left\langle {\boldsymbol{\phi}}_{q}(y) , \widehat{\mathbf{u}}(k_{x,q},k_{z,q};y,t) \right\rangle ,
\end{equation}
since by orthogonality, only the Fourier component at $(k_{x,q}, k_{z,q})$ contributes to the projection. In the MFU180, $\beta_q(t)$ is equivalent to the product of the singular value $\sigma_q$ and right singular vector $\boldsymbol{\psi}_q(t)$, though since the POD is only performed on the MFU180, $\beta_q(t)$ must be calculated by equation \ref{eq:beta} in the FB180 and MFU2200.

\begin{figure}
    \begin{flushleft}
        \hspace{.9cm}
        (a)
        \hspace{6.3cm}
        (b)
    \end{flushleft}
     \centering
     \includegraphics[width=\textwidth,trim={3cm 0 3cm 0},clip]{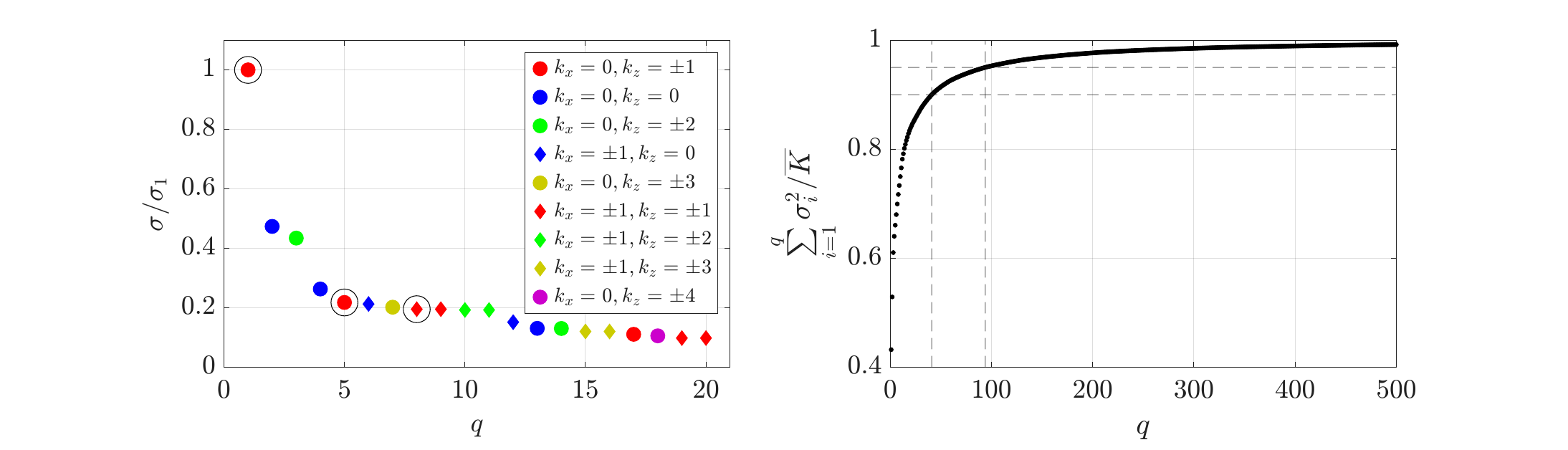}
     
        \caption{(a) Amplification of leading POD modes, normalized by the amplification of the most energetic mode. Circled modes at $q=1$, $5$, and $8$ are the streak, roll-like, and meander modes respectively. (b) Fraction of time-averaged turbulent kinetic energy captured by the first $q$ POD modes. Dashed lines show the number of modes needed to capture 90\% and 95\% of the total energy, which are 41 and 94 respectively.}
        \label{fig:PODmodes}
\end{figure}

In figure \ref{fig:PODmodes}, we show the most energetic POD modes and the portion of the total kinetic energy they capture. From the most energetic POD modes, we focus on those consistent with the SSP in an MFU. The modes $(k_x=0, k_z=\pm1, n=1)$ and $(k_x=0, k_z=\pm1, n=2)$ correspond to streaks and rolls, respectively \citep{waleffe1997self, hamilton1995regeneration}. While streaks are classically associated with streamwise velocity fluctuations and rolls with wall-normal and spanwise motions, we note that the POD modes identified here contain non-negligible energy in all three velocity components, and the streamwise components of both modes account for approximately $96\%$ of their total energy. Being dominated by streamwise fluctuations does not align with the classical view of rolls, so we refer to these modes as roll-like modes. Regardless, these modes capture the dominant spatial structure associated with each SSP component and are sufficient for tracking the evolution of SSP-related energy in time.

We also identify meandering structures of the SSP, which in contrast to streaks and rolls do not have a widely agreed-upon shape, but are known to vary in the streamwise direction ($k_x \neq 0$) and have a strong nonlinear contribution to the roll-like mode. Therefore, we select a POD mode that satisfies these criteria. This nonlinear contribution is quantified using the mode-to-mode energy transfer \citep{ding2025mode}, adapted here for POD modes,
\begin{equation}
\alpha(\ell,p,q;t) = -2\,\mathrm{Re}\left\{ \beta_\ell(t) \beta_p(t) \beta_q(t) \left\langle {\boldsymbol{\phi}}_{q}(y) , \left( {\boldsymbol{\phi}}_p(y) \cdot \nabla \right) {\boldsymbol{\phi}}_\ell(y) \right\rangle \right\},
\end{equation}
which represents the nonlinear energy exchange from mode $\ell$ to mode $q$ via mode $p$ (see appendix~\ref{appB} for the derivation). Here, linear interpolation is used where necessary to ensure that products are evaluated at consistent grid locations. Fixing the recipient to the roll-like mode $(k_{x,q}=0, k_{z,q}=\pm1, n_q=2)$ and sweeping over pairs $(\ell, p)$ with $k_{x,\ell}\neq 0$, we find that the largest transfer occurs with $\ell=(k_x=\pm1,k_z=\pm1,n=1)$ and $p=(k_x=\mp1,k_z=0,n=1)$. We therefore identify $(k_x=\pm1,k_z=\pm1,n=1)$ as the meander mode.

\begin{figure}
    \begin{flushleft}
        \hspace{1.6cm}
        (a)
        \hspace{5.4cm}
        (b)        
    \end{flushleft}
    \centering
    \includegraphics[width=\textwidth]{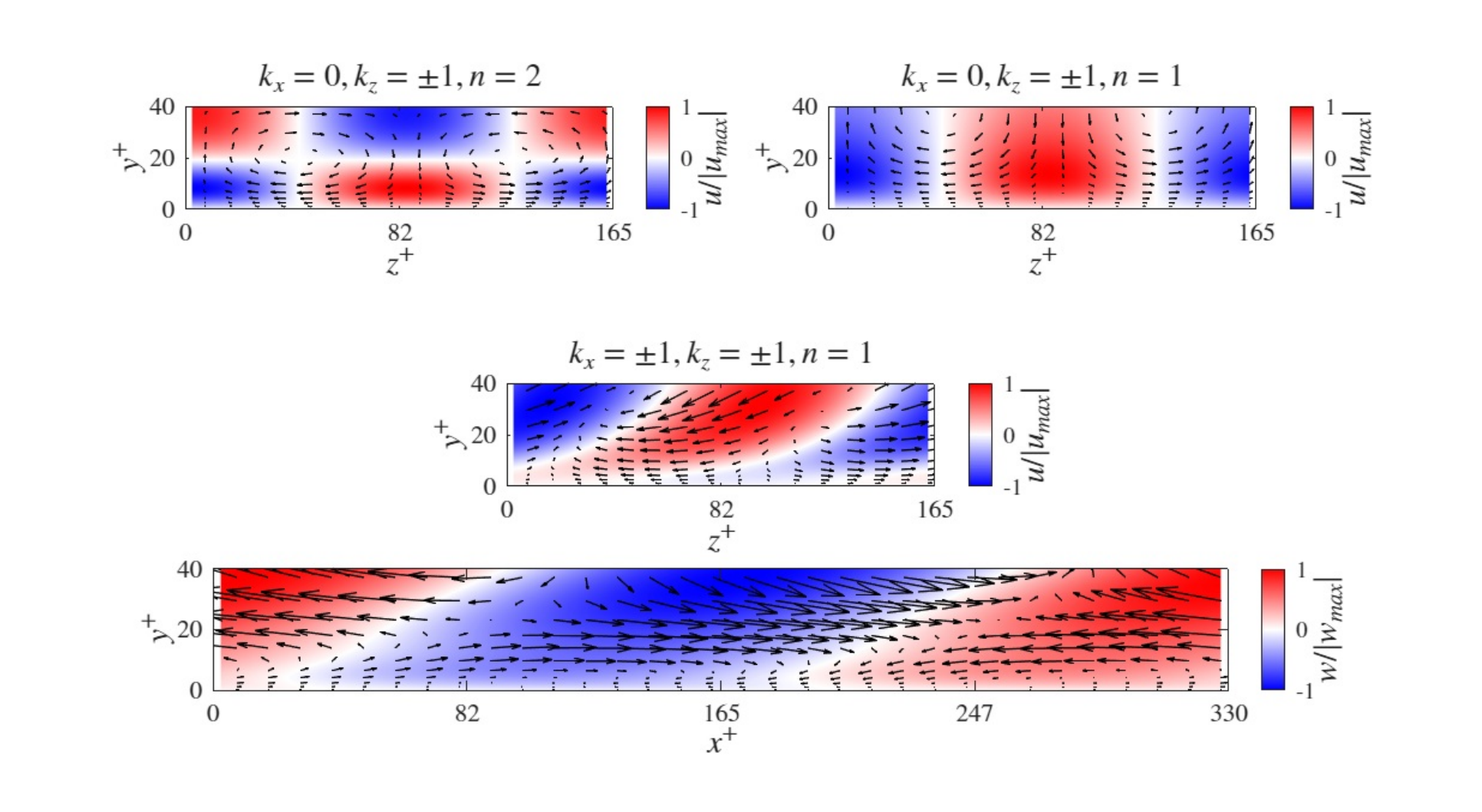}
    \begin{flushleft}
        \vspace{-4.7cm}
        \hspace{1.9cm}
        (c)
        \vspace{4.7cm}
    \end{flushleft}
    \caption{POD modes corresponding to (a) rolls, (b) streaks, and (c) meanders. The streamwise-uniform roll-like and streak modes are shown only in the $y$--$z$ plane, while the meander mode is shown in both the $x$--$y$ and $y$--$z$ planes. For the $y$--$z$ plane ($x$--$y$ plane), color denotes streamwise (spanwise) velocity, and arrows denote spanwise and wall-normal (streamwise and wall-normal) velocities. Velocities are normalized by the maximum velocity component of each mode.}
    \label{fig:POD_SSP}
\end{figure}

The roll-like, streak, and meander modes are denoted $r$, $s$, and $m$, respectively, and visualized in Figure~\ref{fig:POD_SSP}. We track $|\beta_r(t)|$, $|\beta_s(t)|$, and $|\beta_m(t)|$ — the square roots of the energy contained in each mode — over time to follow the evolution of SSP structures.

\subsection{State space representation}
\label{sec:statespace}


To study the relationship between key quantities as the flow evolves over time, we use a state space approach. Variables of interest form axes in a coordinate system, and their instantaneous values define a point in this space. As the flow evolves, this point traces a trajectory through the state space, and the geometry and statistics of this trajectory reveal temporal relationships between variables and dynamical patterns in the flow.

\begin{figure}
     \centering
     \begin{subfigure}[b]{0.45\textwidth}
         \centering
         \caption{}
         \includegraphics[width=\textwidth,trim={0 0 0 1cm},clip]{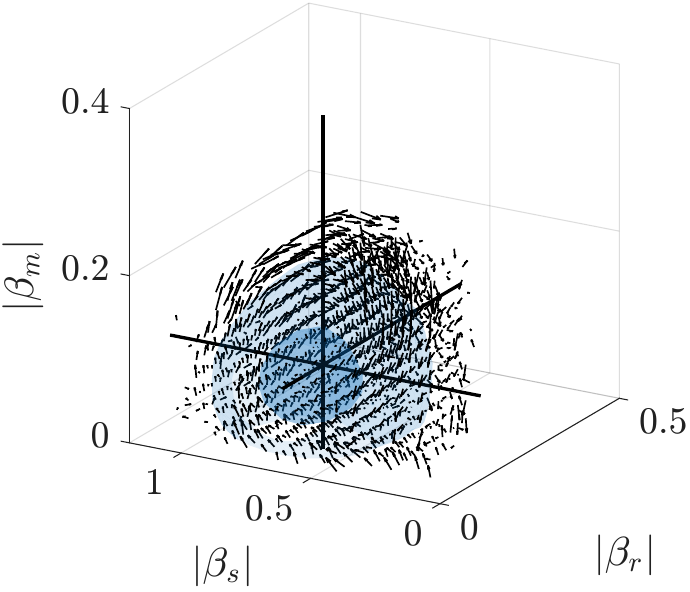}
     \end{subfigure}
     \begin{subfigure}[b]{0.45\textwidth}
         \centering
         \caption{}
         \includegraphics[width=\textwidth,trim={0 0 0 1cm},clip]{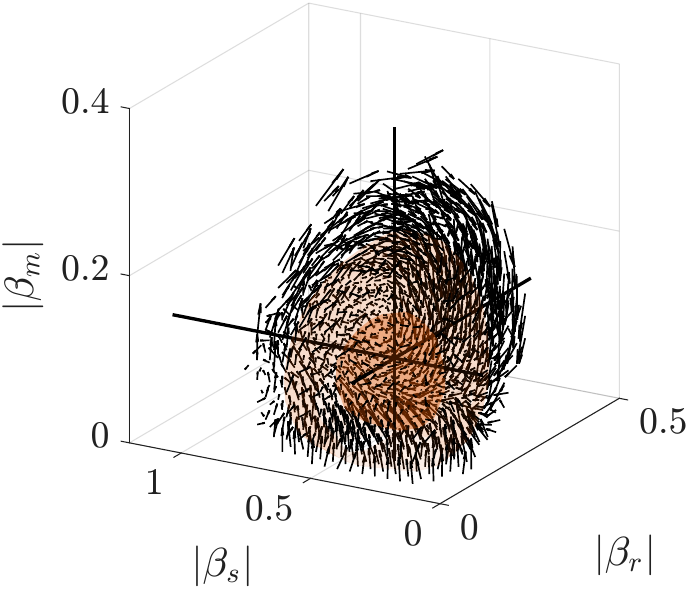}
     \end{subfigure}
        \caption{$(r,s,m)$ state space of the (a) MFU180 and (b)  FB180. Contours contain 75\% and 25\% of occurrences, and arrows show the average speed and direction of trajectory development.}
        \label{fig:mfu_ssp_traj}
\end{figure}

The projection coefficients of the SSP-related modes ($|\beta_r(t)|$, $|\beta_s(t)|$, and $|\beta_m(t)|$) form a three-dimensional state space, as shown in figure \ref{fig:mfu_ssp_traj}. To characterize the trajectory, we first identify the regions of state space most frequently occupied, reflecting the most common contemporaneous combinations of roll-like, streak, and meander mode energies. We also compute the average trajectory by averaging the displacement in state space between consecutive snapshots at each point in the space. The state space is divided into octants based on the median values of $|\beta_r(t)|$, $|\beta_s(t)|$, and $|\beta_m(t)|$, ensuring that the trajectory spends approximately equal time in each octant. The octants are labeled $+$ or $-$ when the quantity is above or below the median. For example, octant $(+r,-s,+m)$ corresponds to above-median roll-like mode energy, below-median mode streak energy, and above-median meander mode energy. Transition probabilities between octants are computed by counting the number of transitions between each pair of octants and dividing by the total number of transitions from the starting octant.

\begin{figure}
     \centering
     \begin{subfigure}[b]{0.99\textwidth}
         \centering
         \includegraphics[width=\textwidth,trim={1cm 4.2cm 1cm 0cm},clip]{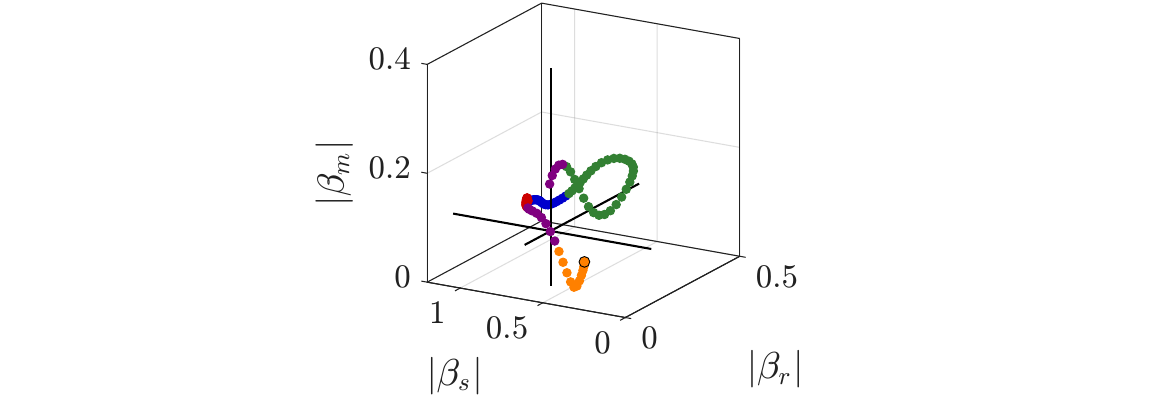}
     \end{subfigure}
     \begin{subfigure}[b]{0.99\textwidth}
         \centering
         \includegraphics[width=\textwidth,trim={0cm 0cm 0cm 9.2cm},clip]{figures/exampletraj_highres.pdf}
     \end{subfigure}
        \caption{Example trajectory through the $(r,s,m)$ state space, beginning at the circled point and progressing over time. State space is divided into octants based on median values of the three quantities, and the color of the trajectory point indicates the different octants. As the flow moves between octants, a sequence of octants is formed, and the time spent in each octant ($T_{seq}^+$) is tracked.}
        \label{fig:exampletraj}
\end{figure}

As the flow moves between octants, a sequence of octants is formed, and the time spent in each octant in the sequence is denoted $T_{seq}^+$. An example sequence is shown in Figure~\ref{fig:exampletraj}. Short visits to an octant may reflect small-scale structures advecting through the sub-domain rather than genuine dynamical processes. 
Based on the distribution of $T^+_{seq}$ across datasets, discussed further in \S ~\ref{sec:temporalanalysis}, we remove sequence entries with $T^+_{seq} \leq 5.3$. We note that small-amplitude trajectories may trace out the same sequence of octants as large-amplitude trajectories, though large-amplitude trajectories reflect more important dynamical processes. However, removing sequence entries with $T^+_{seq} \leq 5.3$ tends to remove very small-amplitude sequence events, and the surviving sequence events have similar average trajectories.

\subsection{Motif analysis}
\label{sec:motif}

To identify patterns in instantaneous dynamics of the SSP-related modes, we apply a network method called motif identification to the octant sequence \citep{milo2002network}. Given a graph, its motifs are small sub-graphs that are statistically overrepresented or underrepresented relative to randomly generated sub-graphs with the same lower-order statistics. Here, lower-order statistics refer to the distribution of sub-sequences shorter than the sub-sequence of interest: to study motifs of length $n$, the randomly generated sequences are constrained to contain the same distribution of sub-sequences of length $n-1$ as the original sequence. In the octant sequence, significant motifs correspond to common paths through state space that occur more frequently than would be expected from pairwise transition statistics alone, thereby capturing dynamical processes rather than time-averaged trends. In this way, motif analysis reveals dynamical patterns that would be missed by simpler statistical analyses. An example of a significant motif is shown in figure~\ref{fig:motif_example}.

\definecolor{red}{rgb}{1,0,0}
\definecolor{green}{rgb}{.1,.8,0}
\definecolor{blue}{rgb}{.2,.3,1}
\definecolor{black}{rgb}{0,0,0}
\begin{figure}
    \centering
    \includegraphics[trim = {4.5cm 1.8cm 3.5cm 0cm},clip,width=\textwidth]{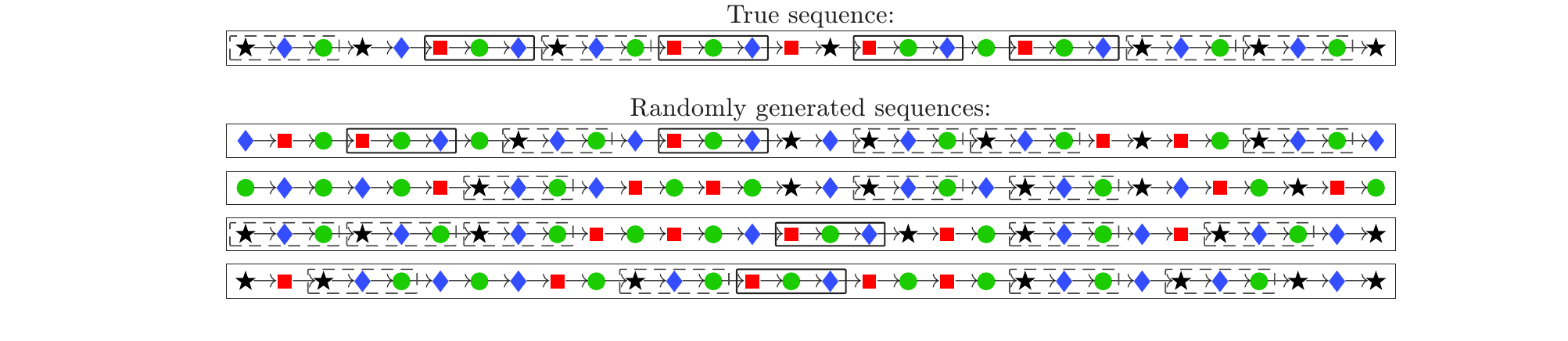}
    \caption{Example of motif significance analysis, showing a true sequence and randomly generated sequences with identical 2-node statistics. For example, because the true sequence contains five instances of $\bigstar \rightarrow \textcolor{blue}{\scalebox{1.25}{$\blacklozenge$}}$, each random sequence is constrained to contain the same number. The motif \textcolor{red}{\scalebox{1.25}{$\blacksquare$}}$\rightarrow$\textcolor{green}{\CIRCLE}$\rightarrow$\textcolor{blue}{\scalebox{1.25}{$\blacklozenge$}} occurs significantly more often in the true sequence than in the random sequences and is therefore identified as a significant motif. In contrast, $\bigstar \rightarrow \textcolor{blue}{\scalebox{1.25}{$\blacklozenge$}} \rightarrow \textcolor{green}{\CIRCLE}$ occurs with comparable frequency in the true and random sequences and is therefore not significant.}
    \label{fig:motif_example}
\end{figure}

We consider sub-sequences comprised of three nodes, where each node is an octant. For each sub-sequence $\mathbf{s}$, a significance score $\widetilde{W}(\mathbf{s})$ is defined as
\begin{equation}
    \widetilde{W}(\mathbf{s})=\frac{f(\mathbf{s})-\overline{f_R(\mathbf{s})}}{ \left( Var(f_R({\mathbf{s}})) \ |\mathbf{S}|/|\mathbf{S}_{SSP}|\right)^{\frac{1}{2}} },
\end{equation}
where $f(\mathbf{s})$ is the frequency of the sub-sequence $\mathbf{s}$ in the true sequence, and  $\overline{f_R(\mathbf{s})}$ is its average frequency over many randomly generated sequences. The significance score measures how much more or less frequently a sub-sequence occurs in the true sequence relative to the random sequences, normalized by the standard deviation ($Var(f_R({\mathbf{s}}))^{1/2}$) of its frequency across the random sequences. Dividing by the standard deviation penalizes sub-sequences whose frequency is highly variable across random sequences, since high variance suggests that the observed frequency could be a sampling artifact rather than a true dynamical pattern; $Var(f_R(\mathbf{s}))$ denotes this variance.

The sequence length $|\mathbf{S}|$ varies between channels due to differences in the amount of available data: $|\mathbf{S}_{\text{MFU180}}|= 1.3 \times 10^4$, $|\mathbf{S}_{\text{FB180}}|= 2.7 \times 10^4$, and $|\mathbf{S}_{\text{MFU2200}}|= 1.4 \times 10^4$. Since the randomly generated sequences match the length of their corresponding true sequence, and the variance of a random sequence is inversely proportional to its length by the law of large numbers, we normalize the variance by $|\mathbf{S}|/|\mathbf{S}_{SSP}|$, where $|\mathbf{S}_{SSP}|$ is the length of a sequence containing one complete cycle of the SSP, which is estimated to be 12 based on the temporal scale of the SSP and the average temporal duration of a sequence event. This normalization makes significance scores comparable across channels with different sequence lengths.





A significance score of zero indicates that a sub-sequence occurs with the same frequency in the true and randomly generated sequences. Since the random sequences are constrained to have the same lower-order statistics as the true sequence, a zero score indicates that the frequency of the sub-sequence is fully explained by its component pairwise transitions rather than as part of a longer dynamical process. For example, if $\bigstar$$\rightarrow$\textcolor{blue}{\scalebox{1.25}{$\blacklozenge$}} and \textcolor{blue}{\scalebox{1.25}{$\blacklozenge$}}$\rightarrow$\textcolor{green}{$\CIRCLE$} both occur frequently in the true sequence, it is not a surprise that $\bigstar$$ \rightarrow $\textcolor{blue}{\scalebox{1.25}{$\blacklozenge$}}$\rightarrow$\textcolor{green}{$\CIRCLE$} also occurs frequently in the true sequence. The random sequences contain the same number of $\bigstar$$\rightarrow $\textcolor{blue}{\scalebox{1.25}{$\blacklozenge$}} and \textcolor{blue}{\scalebox{1.25}{$\blacklozenge$}}$\rightarrow$\textcolor{green}{$\CIRCLE$} as the true sequence, so if the random sequences contain similar frequencies of $\bigstar$$ \rightarrow$\textcolor{blue}{\scalebox{1.25}{$\blacklozenge$}}$\rightarrow$\textcolor{green}{$\CIRCLE$} as the true sequence, its frequency in the true sequence can be entirely attributed to lower-order statistics and it is not a significant motif.

A significance score of one indicates that a sub-sequence occurs one standard deviation more frequently in the true sequence than in the randomly generated sequences, where the standard deviation is computed over sequences that are the size of one SSP cycle, $|\mathbf{S}_{SSP}|$. Because $|\mathbf{S}_{SSP}| \ll |\mathbf{S}|$, the normalization produces relatively small significance scores with magnitudes less than one. Negative scores indicate that a sub-sequence occurs less frequently in the true sequence than in the random sequences.

To generate the random sequences, we use a Markov-chain approach that ensures each random sequence has exactly the same pairwise transition statistics as the true sequence \citep{milo2002network}. Since each node in the sequence contains one incoming edge and one outgoing edge, we employ a ``domino'' approach: the true sequence is split into overlapping 2-node sub-sequences (dominoes), which are then reassembled into a random sequence by drawing dominoes one at a time, requiring that the first node of each new domino matches the last node of the previous one. Recursion is used to ensure that the random sequence does not terminate before all dominoes are used: if remaining dominoes cannot be added to the end of the sequence, they are inserted into an earlier location in the sequence. Compared to simply sampling each node from a fixed distribution, the domino approach guarantees exact preservation of the baseline statistics, so fewer random sequences are needed to produce converged estimates of $\overline{f_R(\mathbf{s})}$ and $Var(f_R(\mathbf{s}))$.

\section{Results}
\label{sec:results}


	




	




In this section, we compare state-space and motif analysis between the three channels. The three channel configurations are designed to isolate progressively more complex dynamical interactions: the MFU180 captures the SSP in the absence of both intra-layer and inter-layer interactions, the FB180 introduces intra-layer interactions between near-wall structures while holding the Reynolds number fixed, and the MFU2200 adds inter-layer interactions between the near-wall region and an energetic outer layer. Due to its restricted domain size, the MFU180 enforces periodicity of the SSP-scale structures, whereas the scale of periodicity in the FB180 and MFU2200 is much larger than the sub-domains of interest, which allows for intra-layer interactions between distinct structures and advection of energy across the sub-domain's boundaries. By comparing state space trajectories and motif significance scores across these configurations, we assess how intra- and inter-layer interactions modify the dynamical patterns of the near-wall region, and whether any patterns persist universally across all three configurations.


\subsection{Effect of intra-layer interactions in the near-wall region}

The MFU180 isolates near-wall coherent structures at the critical scale, i.e. the largest scale the domain supports, whose removal causes relaminarization. This allows the structures' dynamics to develop free of intra-layer interactions and enforces periodicity. The FB180 contains the same energetic scales but accommodates multiple instances of these structures at different spanwise and streamwise locations, so structures tracked within each advecting sub-domain are subject to interactions with neighboring instances of structures of similar size. Comparing the two configurations therefore isolates the effect of intra-layer interactions on near-wall dynamical patterns, while holding the Reynolds number and the relevant viscous-scale structure fixed.


\subsubsection{$(r,s,m)$ state space}
\label{sec:rsmstatespace}

Figure \ref{fig:mfu_ssp_traj} shows the state space of the energy contained in $r$, $s$, and $m$. The average trajectories show a clear cycle from high energy in rolls, to streaks, to meanders, and back to rolls, consistent with the SSP. The same cycle is visible in both the MFU180 and FB180, though the FB180 spans a wider range of values, reflecting higher-amplitude energetic cycles when intra-layer interactions are present. In the low-energy octant $(-r,-s,-m)$, the mean trajectory evolves slowly through a narrow range of state space, while in the high-energy octant $(+r,+s,+m)$ it moves more quickly through a wide range of values, consistent with larger-amplitude structures that develop more rapidly.

\begin{figure}
    \begin{flushleft}
        \hspace{1.9cm}
        (a)
        \hspace{3.1cm}
        (b)      
        \hspace{3.1cm}
        (c)
    \end{flushleft}
     \centering
     \includegraphics[width=\textwidth,trim={2cm 0 3.5cm 2cm},clip]{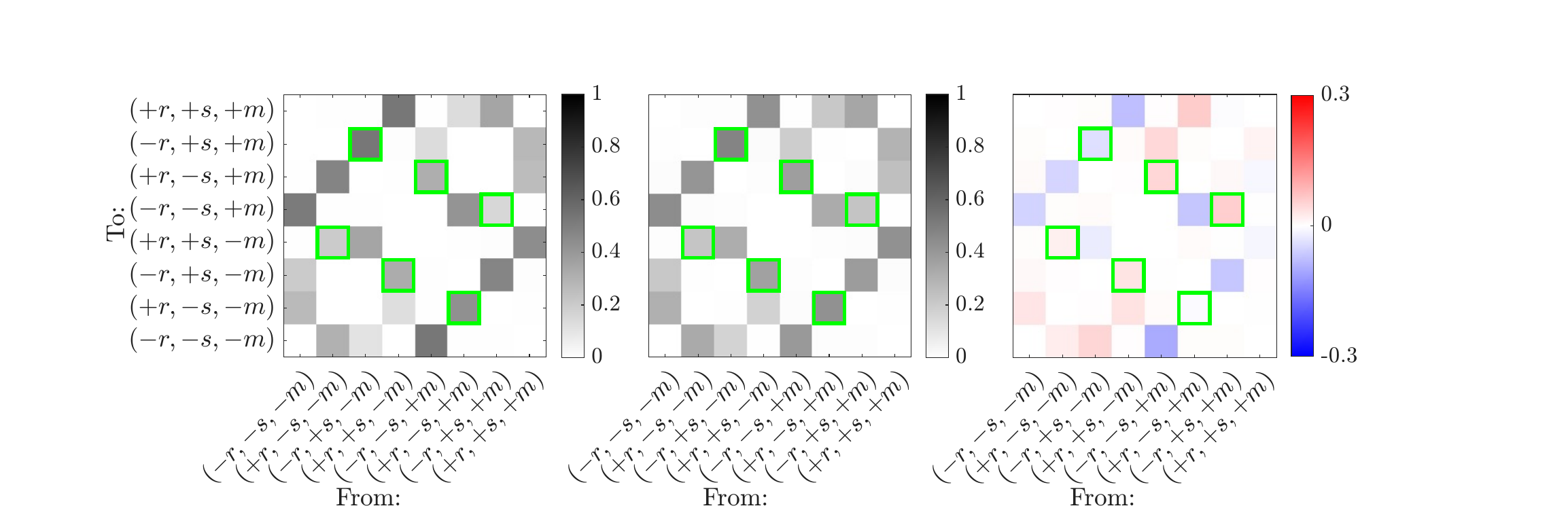}
        \caption{$(r,s,m)$ transition probability for the (a) MFU180 and (b) FB180 from the horizontal axis to the vertical axis. (c) Difference between FB180 and MFU180 transition probabilities, where positive indicates a higher probability in the FB180. Green boxes highlight SSP-related transitions.}
        \label{fig:mfu_ssp_trans}
\end{figure}

Figure \ref{fig:mfu_ssp_trans} shows the transition probabilities of the $(r,s,m)$ state space with SSP-related transitions highlighted in green. Several transitions correspond to steps of the SSP; for example, the presence of high-energy roll-like modes leading to streak amplification corresponds to the $(+r,-s,-m)\to(+r,+s,-m)$ transition. Interestingly, these SSP-related transitions are not the most frequently occurring transitions in the MFU180, despite the mean trajectory being clearly aligned with the SSP cycle. This apparent discrepancy reflects the distinction between time-averaged and instantaneous dynamics: the SSP manifests as a weak statistical drift in state space, visible only when trajectories are averaged over many transitions, but is not a preferred pathway at any given instant. Instantaneously, the flow moves between octants through a broad range of transitions, many of which do not correspond to steps of the SSP. This finding motivates the use of motif analysis, which is designed to identify statistically overrepresented patterns in the instantaneous dynamics rather than time-averaged trends.

The most probable transitions in the MFU180 involve increases in meander mode energy, followed by changes in roll-like mode energy, with streak energy changing least frequently. This ordering reflects the different characteristic timescales of the three structures: meanders are short-lived oblique instabilities that grow and decay rapidly, rolls evolve on intermediate timescales, and streaks are the most coherent and longest-lived of the three, consistent with their role as the primary energy-containing structures in the near-wall region. These differences in temporal dynamics are quantified further in \S\ref{sec:temporalanalysis}. Transitions between diagonally-adjacent octants, which would require simultaneous changes in all three quantities, are rare.

The FB180 transition probabilities, also shown in figure \ref{fig:mfu_ssp_trans}, are qualitatively similar to those of the MFU180, reflecting broadly similar dynamical processes in both channels. SSP-related transitions are again not the most frequent. The differences, shown in figure \ref{fig:mfu_ssp_trans}(c), include a slightly increased presence of diagonal transitions and fewer transitions involving changes in meander energy, both consistent with the faster temporal dynamics of the FB180 identified in \S\ref{sec:temporalanalysis}. Transitions involving changes in roll and streak energy along SSP-related pathways are slightly more common in the FB180, while meander-related SSP transitions are reduced. Whether these shifts reflect genuine dynamical differences due to intra-layer interactions, or are partially an artifact of energetic structures drifting across sub-domain boundaries, cannot be resolved from transition probabilities alone. Motif analysis, which identifies statistically overrepresented sequences of transitions rather than individual transition frequencies, is better suited to isolating genuine dynamical differences between the two channels.

\subsubsection{$(r,s,m)$ network motifs}

\begin{figure}
    \centering
    \includegraphics[width=0.8\textwidth]{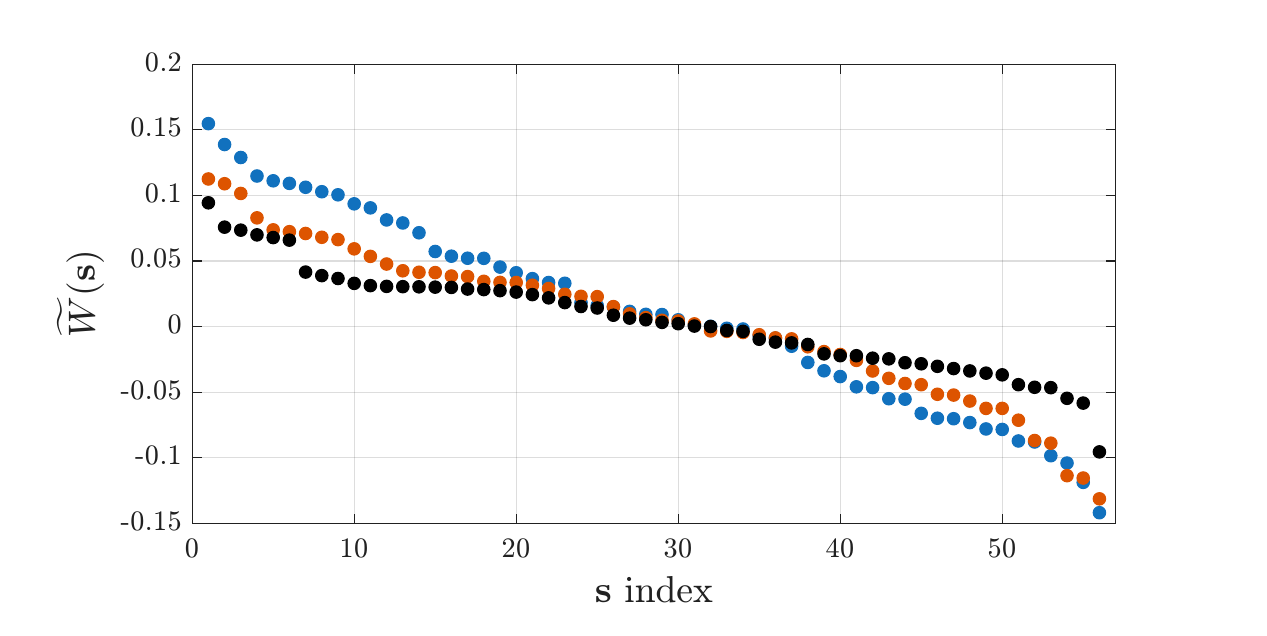}
    \caption{Significance scores of sub-sequences of length three, sorted highest to lowest, for the MFU180 (blue), FB180 (red), and MFU2200 (black). For visualization purposes, the 56 most frequent sub-sequences are plotted, since this corresponds to the number of subsequences with $f(\mathbf{s})>0.005$ in the MFU180.}
    \label{fig:sigscoredist}
\end{figure}

Next, significance scores for all octant sub-sequences of length three are computed. Sub-sequences that occur very infrequently in the true sequence (for example, sub-sequences that involve transitions between diagonally-adjacent octants) have low standard deviations in the randomly generated sequences, resulting in high significance scores. These low frequency sub-sequences are unlikely to be important to the overall dynamics, so we do not report the significance scores of sub-sequences with frequencies $f(\mathbf{s})\leq0.005$. The cutoff selection is discussed in appendix \ref{appC}.

Figure \ref{fig:sigscoredist} shows the significance scores of sub-sequences of length three, ranked highest to lowest. Most sub-sequences follow a roughly linear trend,  though a small number have very high or very low scores, suggesting that only a few dynamical patterns are particularly important to the near-wall dynamics. These outlying patterns occur significantly more or less frequently than would be predicted by their lower-order statistics, which demonstrates the advantage of motif analysis over simpler statistical tools. To ensure a sufficient number of randomly generated sequences were used to compute $\overline{f_R(\mathbf{s})}$ and $Var(f_R({\mathbf{s}}))$, we perform a sensitivity study (see appendix~\ref{appC}).

The overall distribution of $\widetilde{W}(\mathbf{s})$ is similar across the three datasets, but the MFU180 has larger magnitude scores at both extremes. This reflects the fact that in the MFU180, the sub-domain captures the full spatial extent of the turbulence, so the key dynamical processes sustaining turbulence must occur repeatedly within it, producing strongly overrepresented motifs. In the FB180 and MFU2200, by contrast, structures inside the sub-domain are influenced by neighboring near-wall structures and outer-layer dynamics respectively, so no single dynamical process is required to occur as frequently to sustain turbulence, and significance scores are correspondingly reduced.

\begin{table}
\begin{center}
\renewcommand{\arraystretch}{1} 
\begin{tabular}{c|c||c|c}
  $s$ & $\widetilde{W}(s)$ & $s$ & $\widetilde{W}(s)$\\
\scriptsize${(+r,+s,-m)\to(+r,+s,+m)\to(-r,+s,+m)}$&0.1551& \scriptsize${(-r,+s,+m)\to(+r,+s,+m)\to(-r,+s,+m)}$&-0.1423 \\
\textcolor{blue}{\scriptsize${(-r,-s,+m)\to(+r,-s,+m)\to(+r,-s,-m)}$}&0.1390& \scriptsize${(-r,+s,-m)\to(+r,+s,-m)\to(-r,+s,-m)}$&-0.1187 \\
\scriptsize${(-r,+s,+m)\to(+r,+s,+m)\to(+r,-s,+m)}$&0.1288& \scriptsize${(+r,+s,-m)\to(-r,+s,-m)\to(+r,+s,-m)}$&-0.1037 \\
\scriptsize${(-r,-s,+m)\to(-r,+s,+m)\to(-r,+s,-m)}$&0.1146& \scriptsize${(-r,+s,-m)\to(-r,+s,+m)\to(-r,+s,-m)}$&-0.0983 \\
\scriptsize${(-r,+s,+m)\to(-r,+s,-m)\to(+r,+s,-m)}$&0.1108& \scriptsize${(+r,-s,-m)\to(+r,-s,+m)\to(+r,-s,-m)}$&-0.0884 
\end{tabular}
\captionof{table}{(left) Five most significant and (right) five least significant motifs of length three for the MFU180. Blue motifs are related to the SSP.}
\label{tab:motifs_mfu}
\end{center}
\end{table}

\begin{table}
\begin{center}
\renewcommand{\arraystretch}{1} 
\begin{tabular}{c|c||c|c}
  $s$ & $\widetilde{W}(s)$ & $s$ & $\widetilde{W}(s)$\\
\scriptsize${(+r,-s,-m)\to(+r,-s,+m)\to(-r,-s,+m)}$&0.1127 & \scriptsize${(+r,-s,-m)\to(+r,-s,+m)\to(+r,-s,-m)}$&-0.1160\\
\scriptsize${(-r,+s,+m)\to(+r,+s,+m)\to(+r,-s,+m)}$&0.0876 & \scriptsize${(+r,-s,+m)\to(-r,-s,+m)\to(+r,-s,+m)}$&-0.1049 \\
\scriptsize${(+r,+s,-m)\to(+r,+s,+m)\to(-r,+s,+m)}$&0.0864 & \scriptsize${(-r,+s,+m)\to(+r,+s,+m)\to(-r,+s,+m)}$&-0.1047 \\
\scriptsize${(+r,-s,+m)\to(-r,-s,+m)\to(-r,-s,-m)}$&0.0788 & \scriptsize${(-r,-s,-m)\to(+r,-s,-m)\to(-r,-s,-m)}$&-0.0787 \\
\textcolor{blue}{\scriptsize${(-r,-s,-m)\to(-r,-s,+m)\to(+r,-s,+m)}$}&0.0704 & \scriptsize${(-r,-s,-m)\to(-r,-s,+m)\to(-r,-s,-m)}$&-0.0680
\end{tabular}
\captionof{table}{(left) Five most significant and (right) five least significant motifs of length three for the FB180. Blue motifs are related to the SSP.}
\label{tab:motifs_fc}
\end{center}
\end{table}

Table \ref{tab:motifs_mfu} shows the five most and least significant motifs in the MFU180. The second-most significant motif, $(-r,-s,+m) \to (+r,-s,+m) \to (+r,-s,-m)$, captures meander mode energy peaking, roll-like mode energy rising, and meander mode energy subsequently decaying, consistent with the SSP step in which meanders transfer energy to rolls. However, the remaining four most significant motifs do not correspond to SSP steps. The most significant motif, $(+r,+s,-m) \to (+r,+s,+m) \to (-r,+s,+m)$, and the third most significant, $(-r,+s,+m) \to (+r,+s,+m) \to (+r,-s,+m)$, both involve the overall high-energy state $(+r,+s,+m)$ and describe specific pathways by which the flow enters and leaves it, suggestive of intermittent burst events. The fourth and fifth most significant motifs, $(-r,-s,+m) \to (-r,+s,+m) \to (-r,+s,-m)$ and $(-r,+s,+m) \to (-r,+s,-m) \to (+r,+s,-m)$, form an overlapping sequence suggesting a longer pattern of energy redistribution among the three modes. That several non-SSP motifs are among the most significant in the MFU180 suggests that dynamical processes beyond the SSP play an important role in sustaining near-wall turbulence even in the most idealized configuration.


Table \ref{tab:motifs_fc} shows the five most and least significant motifs in the FB180. The fifth most significant motif, $(-r,-s,-m) \to (-r,-s,+m) \to (+r,-s,+m)$, captures meander mode energy rising followed by roll-like mode energy rising, consistent with the SSP step in which meanders transfer energy to rolls, though the specific sequence differs from the analogous MFU180 motif. The most significant FB180 motif, $(+r,-s,-m) \to (+r,-s,+m) \to (-r,-s,+m)$, and the motif $(-r,+s,+m) \to (+r,+s,+m) \to (+r,-s,+m)$ are also significant in the MFU180, indicating shared dynamical processes between the two channels. The presence of these common motifs, including both SSP-related and burst-related patterns, supports the idea that the dominant near-wall dynamical processes are broadly similar across the two configurations despite the presence of intra-layer interactions in the FB180. This suggests a degree of universality in near-wall dynamics that extends beyond the SSP itself.



Combinations of motifs with overlapping octants suggest the presence of longer dynamical patterns involving sequences of more than three octants. In the MFU180, the fourth and fifth most significant motifs are part of the pattern $(-r,-s,+m)\to(-r,+s,+m)\to(-r,+s,-m)\to(+r,+s,-m)$, and in the FB180, the first and fourth most significant motifs combine into $(+r,-s,-m)\to(+r,-s,+m)\to(-r,-s,+m)\to(-r,-s,-m)$. Extending this analysis to sub-sequences of length four would likely reveal further insights into these longer patterns, though this would require substantially more data and is left for future work.

For both the MFU180 and FB180, all five least significant motifs are back-and-forth transitions that return to the octant they started in.  The under-representation of these patterns in the true sequences indicates that near-wall turbulence does not sustain itself through rapid oscillation between pairs of states, but instead progresses through longer sequences of distinct dynamical configurations. This is consistent with a sustaining mechanism that relies on continued energy exchange among rolls, streaks, and meanders across a range of states rather than repeated back-and-forth cycling between two.

\subsection{Effect of outer layer on near-wall dynamics}

	




The MFU2200 introduces inter-layer interactions between the near-wall region and an energetic outer layer. Like the FB180, its near-wall region contains multiple instances of the critical-scale structures, so that structures tracked within each advecting sub-domain are subject to intra-layer interactions. Comparing the MFU2200 with the FB180 therefore isolates the additional effect of inter-layer interactions on near-wall dynamical patterns. We further examine this effect by conditioning the near-wall results on whether the outer-layer turbulent kinetic energy $K_{OL}$ is above or below its mean value, allowing us to distinguish the near-wall dynamics associated with a quiescent versus an energetic outer layer. There are an equal number of above-mean $K_{OL}$ and below-mean $K_{OL}$ trajectories.

\begin{figure}
    \begin{flushleft}
        \hspace{.8cm}
        (a)
        \hspace{5cm}
        (b)      
        \hspace{2.9cm}
        (c)
    \end{flushleft}
     \centering
     \begin{subfigure}[b]{0.32\textwidth}
         \centering
         \includegraphics[width=\textwidth,trim={0cm 0cm .5cm 0cm},clip]{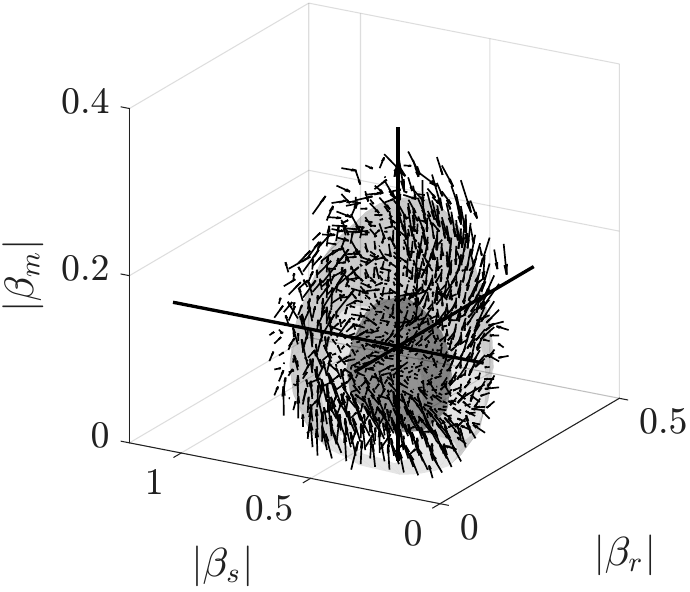}
     \end{subfigure}
     \begin{subfigure}[b]{0.67\textwidth}
         \centering
         \includegraphics[width=\textwidth,trim={0cm 0cm 3cm 2cm},clip]{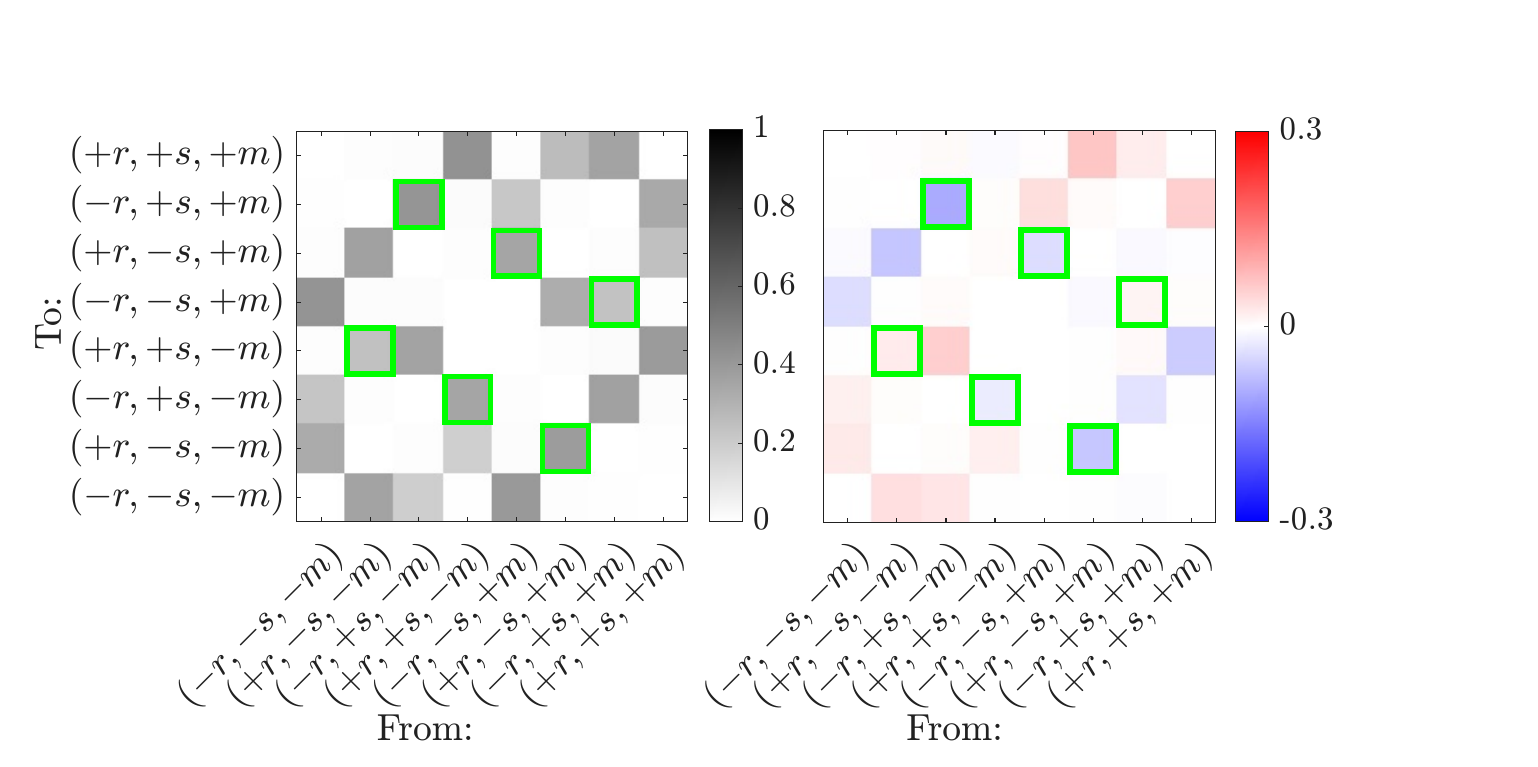}
     \end{subfigure}
        \caption{(a) $(r,s,m)$ state space of the MFU2200. Contours contain 75\% and 25\% of occurrences, and arrows show the average speed and direction of trajectory development. (b) SSP-mode transition probability for the MFU2200 from the horizontal axis to the vertical axis. (c) Difference between MFU2200 and FB180 transition probabilities, where positive indicates a higher probability in the MFU2200. Green boxes highlight SSP-related transitions. }
        \label{fig:llmfu_ssp_traj}
\end{figure}

Figure \ref{fig:llmfu_ssp_traj}(a) shows the $(r,s,m)$ state of the MFU2200, where the SSP cycle is visible in the mean trajectory but is more chaotic and spans a wider range of values than in the MFU180 and FB180. Figure \ref{fig:llmfu_ssp_traj}(b) and (c) show the transition probability and its discrepancy with the FB180, respectively. Generally, the transition probabilities are similar to the FB180, though transitions involving changes in meander energy are slightly less common. This reflects the broader range of dynamical pathways available when inter-layer interactions are present: rather than being funneled through a limited set of processes as in the MFU180 and FB180, the near-wall structures of the MFU2200 can develop through a variety of different interactions involving both near-wall and outer-layer structures.

Motif analysis of the unconditioned MFU2200 reveals SSP-related patterns among both the most and least significant motifs. The two least significant SSP-related motifs, $(-r,-s,-m)\to(-r,+s,-m)\to(-r,+s,+m)$ and $(+r,-s,-m)\to(+r,+s,-m)\to(+r,+s,+m)$ both correspond to the streak-to-meander step of the SSP, indicating that this pathway is underrepresented relative to what pairwise transition statistics alone would predict. This suggests that in the MFU2200, energy transfer from streamwise-constant to streamwise-varying modes occurs preferentially through other mechanisms, such as the streak instability pathway described by \cite{schoppa2002coherent}, rather than through the canonical SSP sequence.

The most significant motif, $(-r,+s,+m) \to (+r,+s,+m) \to (+r,-s,+m)$, is shared across all three channels, highlighting a dynamical process that persists regardless of whether intra- or inter-layer interactions are present. This motif describes roll-like mode energy rising to an overall high-energy state followed by streak mode energy decaying, consistent with a streak bursting event. Four of the five most significant motifs contain the high energy state $(+r,+s,+m)$, compared with two each in the MFU180 and FB180, indicating that patterned entry into and exit from this high-energy state is a more prominent feature of the dynamics at higher Reynolds numbers. The flow does not reach this bursting state randomly but follows specific dynamics through state space, and the regularity of these sequences become more pronounced when inter-layer interactions are present. 

Among the least significant motifs, only two are back-and-forth transitions, compared with all five in the MFU180 and FB180. Combined with the underrepresentation of SSP-related pathways, this suggests that when inter-layer interactions are present, the near-wall region does not need to sustain turbulence independently through a fixed set of dynamical processes. Instead, energy supplied by outer-layer structures provides an additional sustaining mechanism, reducing the reliance on any particular near-wall pathway and allowing occasional back-and-forth transitions without compromising turbulence maintenance.

\begin{table}
\begin{center}
\renewcommand{\arraystretch}{1} 
\begin{tabular}{c|c||c|c}
  $s$ & $\widetilde{W}(s)$ & $s$ & $\widetilde{W}(s)$ \\
  \hline   
\scriptsize${(-r,+s,+m)\to(+r,+s,+m)\to(+r,-s,+m)}$&0.0950 & \scriptsize${(-r,+s,+m)\to(+r,+s,+m)\to(-r,+s,+m)}$&-0.0961\\
\textcolor{blue}{\scriptsize${(+r,+s,-m)\to(-r,+s,-m)\to(-r,+s,+m)}$}&0.0765 & \textcolor{blue}{\scriptsize${(-r,-s,-m)\to(-r,+s,-m)\to(-r,+s,+m)}$}&-0.0589\\
\scriptsize${(-r,-s,+m)\to(-r,+s,+m)\to(+r,+s,+m)}$&0.0728 & \textcolor{blue}{\scriptsize${(+r,-s,-m)\to(+r,+s,-m)\to(+r,+s,+m)}$}&-0.0542 \\
\textcolor{blue}{\scriptsize${(+r,+s,+m)\to(+r,+s,-m)\to(-r,+s,-m)}$}&0.0704 & \scriptsize${(+r,+s,-m)\to(+r,+s,+m)\to(-r,+s,+m)}$&-0.0469 \\
\scriptsize${(+r,+s,-m)\to(+r,+s,+m)\to(-r,+s,-m)}$&0.0677 & \scriptsize${(+r,+s,+m)\to(-r,+s,+m)\to(+r,+s,+m)}$&-0.0462
\vspace{5pt}
\end{tabular}
\captionof{table}{(left) Five most significant and (right) five least significant motifs of length three for the MFU2200. Blue motifs are related to the SSP.}
\label{tab:motifs}
\end{center}
\end{table}


\subsubsection{Conditioning on outer-layer energy}

\begin{figure}
    \begin{flushleft}
        \hspace{1cm}
        (a)
        \hspace{3.8cm}
        (b)      
        \hspace{4.5cm}
        (c)
    \end{flushleft}
     \centering
     \begin{subfigure}[b]{0.32\textwidth}
         \centering
         \includegraphics[width=\textwidth,trim={0cm 0cm .5cm 0cm},clip]{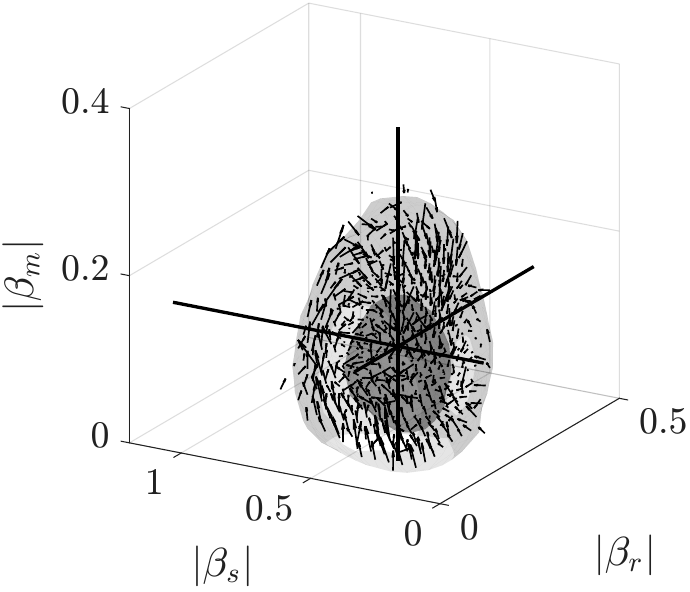}
     \end{subfigure}
     \begin{subfigure}[b]{0.32\textwidth}
         \centering
         \includegraphics[width=\textwidth,trim={0cm 0cm .5cm 0cm},clip]{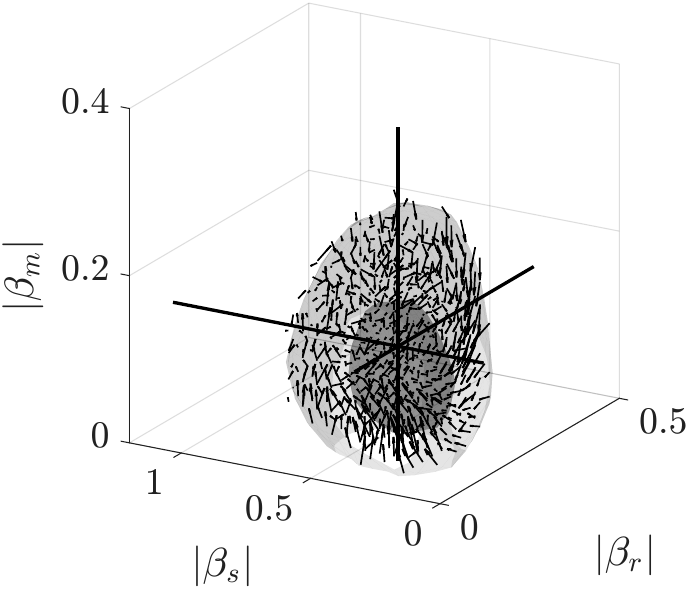}
     \end{subfigure}
     \begin{subfigure}[b]{0.32\textwidth}
         \centering
         \includegraphics[width=\textwidth,trim={0cm 0cm .6cm 1.4cm},clip]{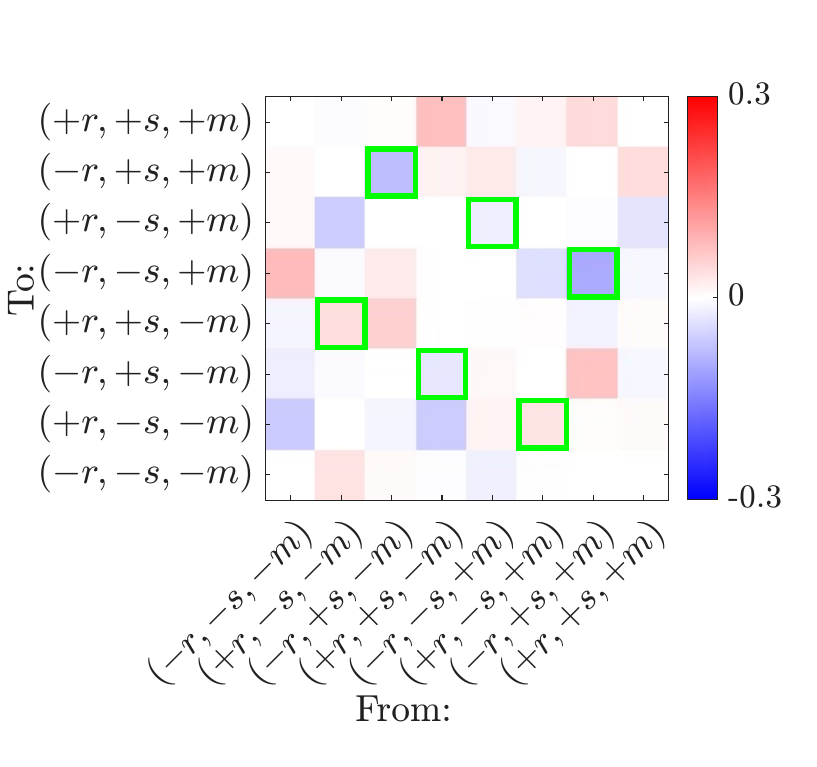}
     \end{subfigure}
        \caption{$(r,s,m)$ state space of the conditioned MFU2200, when (a) $K_{OL}$ is below its mean, and (b) when $K_{OL}$ is above its mean. Contours contain 75\% and 25\% of occurrences, and arrows show the average speed and direction of trajectory development. (c) Difference between the high and low $K_{OL}$ transition probabilities, where positive indicates a higher probability in the high $K_{OL}$. Green boxes highlight SSP-related transitions.}
        \label{fig:llmfu_ssp_traj_cond}
\end{figure}

Figure \ref{fig:llmfu_ssp_traj_cond} shows the $(r, s, m)$ state space conditioned on $K_{OL}$ being above and below its mean value. The mean trajectories are visually similar in both cases: the SSP cycle is not clearly visible in either, and the trajectories span comparable regions of state space. Amplitude modulation by outer-layer structures would be expected to produce differences in the mean trajectory, since large-scale structures are known to modulate the amplitude of near-wall fluctuations \cite{hutchins2007evidence, mathis2009large}. However, amplitude modulation either increases or decreases near-wall energy depending on the phase relationship between the inner and outer layers, so its positive and negative effects may cancel when averaged over all trajectories at a given $K_{OL}$ level, leaving the mean trajectories similar despite the differing outer-layer states. Interestingly, the SSP appears more visible in the unconditioned data (\ref{fig:llmfu_ssp_traj}(a)) than in either of the conditioned datasets; this may be because the unconditioned dataset contains twice the amount of snapshots as the conditioned datasets, so its average trajectory is more converged toward the SSP.

More pronounced differences emerge in the transition probability discrepancy between the high and low $K_{OL}$ cases, shown in figure \ref{fig:llmfu_ssp_traj_cond}(c). SSP-related transitions, highlighted in green, tend to occur less frequently when $K_{OL}$ is high. This is consistent with the interpretation that inter-layer interactions provide an additional sustaining mechanism for near-wall turbulence: when the outer-layer is energetic, near-wall dynamics are less reliant on the SSP to maintain themselves, and SSP-related transitions are correspondingly depressed. When $K_{OL}$ is low, this external support is reduced, and the near-wall region must sustain turbulence more autonomously, producing a stronger presence of the SSP-related transitions.

Table \ref{tab:motifs2} shows that the high and low $K_{OL}$ cases share no common significant motifs, revealing that fundamentally different dynamical processes drive near-wall turbulence depending on the energy state of the outer layer. When $K_{OL}$ is below its mean, the most significant motifs overlap substantially with those of the FB180 and MFU180. The most significant low-$K_{OL}$ motif, $(-r,+s,+m) \to (+r,+s,+m) \to (+r,-s,+m)$, is significant in both the MFU180 and FB180, and the motifs $(+r,-s,-m) \to (+r,-s,+m) \to (-r,-s,+m)$ and $(+r,-s,+m) \to (-r,-s,+m) \to (-r,-s,-m)$ are both significant in the FB180, forming part of the longer four-node pattern $(+r,-s,-m) \to (+r,-s,+m) \to (-r,-s,+m) \to (-r,-s,-m)$. The least significant low-$K_{OL}$ motifs are all back-and-forth transitions, consistent with the MFU180 and FB180. Taken together, these similarities suggest that when the outer layer is quiescent, near-wall dynamics resemble those of a lower Reynolds number flow without a distinct outer layer, sustaining turbulence through intra-layer interactions among near-wall structures alone.

When $K_{OL}$ is above its mean, the significant motifs are substantially different from those of the MFU180 and FB180. Only $(-r,-s,-m) \to (-r,-s,+m) \to (+r,-s,+m)$ is shared with the FB180, and none are shared with the MFU180. Four of the five most significant motifs contain the high-energy octant $(+r,+s,+m)$, compared with only one in the low $K_{OL}$ case, reflecting the prominence of intermittent bursting events driven by outer-layer forcing when $K_{OL}$ is high. The flow follows regular pathways to enter and leave this high-energy state, suggesting that burst events are not random but are structured by the inter-layer interactions. Among the least significant motifs, two are back-and-forth transitions, in contrast to the low $K_{OL}$ case where all five are back-and-forth. Rather than indicating simple oscillation between states, these back-and-forth transitions under high $K_{OL}$ reflect the disruptive effect of energetic outer-layer structures on near-wall trajectories, occasionally driving the flow back to a previous state before it can progress further through state space. That near-wall turbulence can still be sustained under these conditions points to the outer layer itself as an additional energy source that compensates for the disruption of near-wall dynamical sequences.


\begin{table}
\begin{center}
\renewcommand{\arraystretch}{1} 
\begin{tabular}{c|c||c|c}
  $s$ & $\widetilde{W}(s)$ & $s$ & $\widetilde{W}(s)$  \\
  \hline   
\scriptsize${(-r,+s,+m)\to(+r,+s,+m)\to(+r,-s,+m)}$&0.1585 & \scriptsize${(+r,+s,-m)\to(+r,+s,+m)\to(+r,+s,-m)}$&-0.1047\\
\scriptsize${(+r,-s,-m)\to(+r,-s,+m)\to(-r,-s,+m)}$&0.1265 & \scriptsize${(-r,+s,+m)\to(+r,+s,+m)\to(-r,+s,+m)}$&-0.1012\\
\scriptsize${(-r,-s,+m)\to(-r,-s,-m)\to(+r,-s,-m)}$&0.1158 & \scriptsize${(-r,-s,+m)\to(-r,-s,-m)\to(-r,-s,+m)}$&-0.0843\\
\scriptsize${(+r,-s,+m)\to(-r,-s,+m)\to(-r,-s,-m)}$&0.1085 & \scriptsize${(-r,+s,-m)\to(+r,+s,-m)\to(-r,+s,-m)}$&-0.0825 \\
\textcolor{blue}{\scriptsize${(-r,+s,+m)\to(-r,-s,+m)\to(+r,-s,+m)}$}&0.0998 & \scriptsize${(+r,+s,+m)\to(-r,+s,+m)\to(+r,+s,+m)}$&-0.0750 \\
\hline
  $s$ & $\widetilde{W}(s)$ & $s$ & $\widetilde{W}(s)$\\
  \hline   
\scriptsize${(+r,+s,+m)\to(-r,+s,+m)\to(+r,+s,-m)}$&0.1907 & \scriptsize${(-r,-s,-m)\to(+r,-s,-m)\to(+r,-s,+m)}$&-0.1320 \\
\textcolor{blue}{\scriptsize${(-r,+s,+m)\to(+r,+s,+m)\to(+r,+s,-m)}$}&0.1231 & \scriptsize${(-r,+s,+m)\to(+r,+s,+m)\to(-r,+s,+m)}$&-0.0910\\
\textcolor{blue}{\scriptsize${(-r,-s,-m)\to(-r,-s,+m)\to(+r,-s,+m)}$}&0.1167 & \textcolor{blue}{\scriptsize${(-r,-s,-m)\to(-r,+s,-m)\to(-r,+s,+m)}$}&-0.0872 \\
\scriptsize${(-r,-s,+m)\to(-r,+s,+m)\to(+r,+s,+m)}$&0.1128 & \scriptsize${(+r,-s,-m)\to(-r,-s,-m)\to(+r,-s,-m)}$&-0.0775\\
\scriptsize${(-r,+s,-m)\to(+r,+s,+m)\to(-r,+s,+m)}$&0.1079 & \scriptsize${(+r,+s,-m)\to(+r,+s,+m)\to(+r,-s,+m)}$&-0.0709 
\end{tabular}
\vspace{5pt}
\captionof{table}{(left) Five most significant and (right) five least significant motifs of length three for the MFU2200, conditioned to when the $K_{OL}$ is (upper) below its mean value and (lower) above its mean value. Blue motifs are related to the SSP.}
\label{tab:motifs2}
\end{center}
\end{table}

\subsection{Temporal scales of dynamics} 
\label{sec:temporalanalysis}

\begin{figure}
    \begin{subfigure}[b]{0.52\textwidth}
     \centering
     \caption{}
     \vspace{-.4cm}
    \includegraphics[width=\textwidth,trim={.8cm 0cm 2.3cm 0cm},clip]{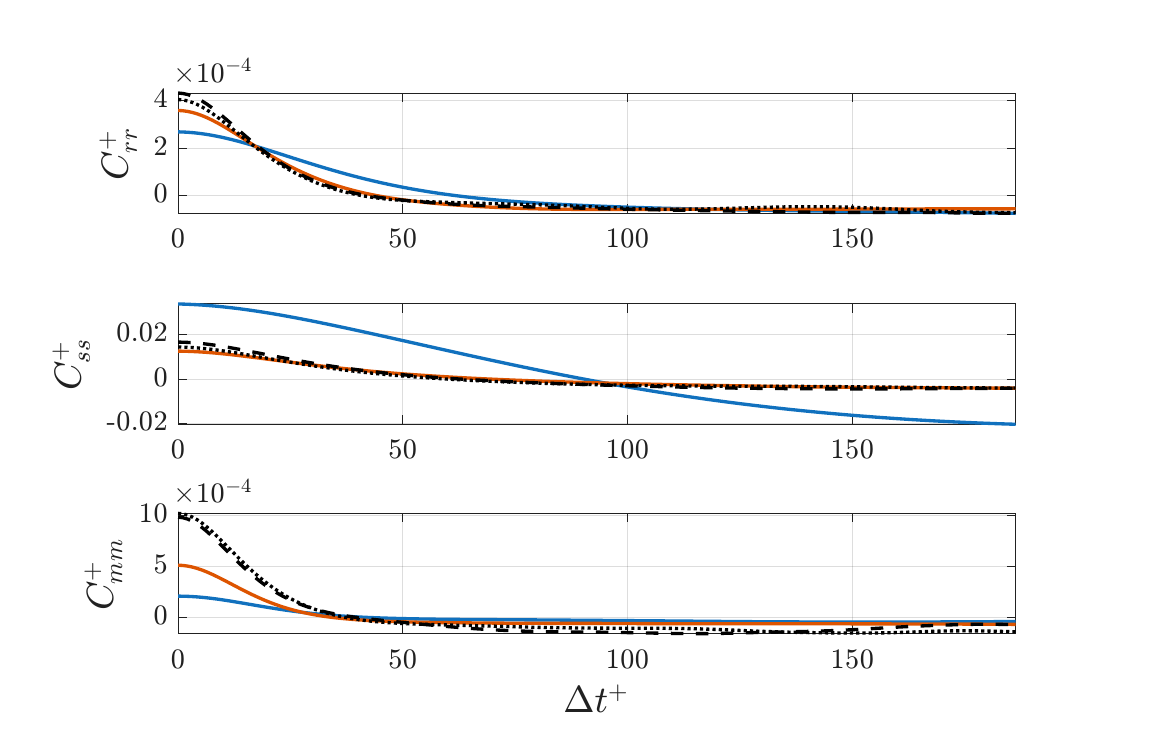}
    \end{subfigure}
     \begin{subfigure}[b]{0.47\textwidth}
     \centering
     \vspace{-.1cm}
     \caption{}
     \vspace{-.2cm}
    \includegraphics[width=\textwidth,trim={.2cm 0cm 2.1cm 0cm},clip]{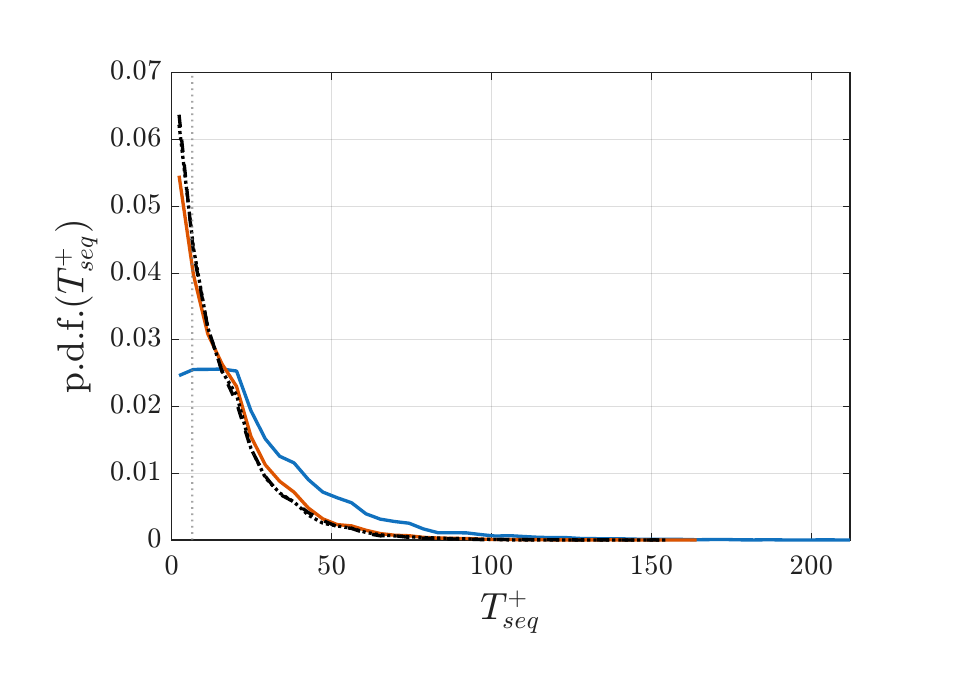}
    \end{subfigure}
        \caption{(a) Auto-covariance of projection coefficient of $r$, $s$, and $m$ for the MFU180 (blue), FB180 (red), and MFU2200 with low $K_{OL}$ (black dotted) and high $K_{OL}$ (black dashed). (b) Histogram of time spent in each octant for the same datasets. Grey dotted line indicates the cutoff value of $T_{seq}^+ = 5.3$}
        \label{fig:traj_histogram}
\end{figure}

The motif significance scores and transition probabilities reported in the previous sections reflect not only the structure of dynamical pathways but are also indirectly affected by the timescales on which they operate. Differences in temporal scales between the three channels therefore provide important context for interpreting the motif results. Figure \ref{fig:traj_histogram}(a) shows the temporal autocovariance of $|\beta_r(t)|$, $|\beta_s(t)|$, and $|\beta_m(t)|$ up to a time lag of $\Delta t^+=180$, corresponding to one eddy turnover time in the $Re_\tau \approx 180$ channels. Rather than a normalized correlation, the raw covariance is shown to emphasize the difference in magnitude between the structures. The streak mode has a much larger covariance magnitude than the roll-like or meander modes, particularly in the MFU180, reflecting their stronger coherence, and they also decay the most slowly, confirming that streaks are the longest-lived of the three structures. The meander modes decay most rapidly, consistent with their role as short-lived oblique instabilities.
Structures in the MFU180 have longer lifetimes overall, with covariances reaching zero after a much longer time lag than the other two channels, indicating that near-wall dynamics in the MFU180 are slower due to the absence of intra-layer interactions. This trend is most pronounced for the streak mode and least pronounced for the meander mode. 
Interestingly, while most structures asymptote to a small negative covariance, the streak mode in the MFU180 continues to a high negative covariance, suggesting a strong cycle of high and low energy streaks over temporal scales longer than $\Delta t^+=186$. 

These slower dynamics provide additional context for the larger motif significance scores observed in the MFU180 relative to the FB180 and MFU2200. The lack of intra- and inter-layer interactions in the MFU180 limit the accessible dynamical pathways and slow the speed at which energy can be transferred between structures. As a result, the MFU180 undergoes slower dynamical processes which repeat frequently, producing more statistically pronounced motifs.



Figure~\ref{fig:traj_histogram}(b) shows the distribution of $T^+_{seq}$ across datasets. The MFU180 has few very short sequence events and a plateau at small $T^+_{seq}$, consistent with its slower dynamics. The FB180 and MFU2200 have many very short events, likely related to structures moving in and out of the sub-domains rather than genuine dynamical transitions, which motivated the removal of sequence entries with $T^+_{seq} \leq 5.3$ for all three datasets. Removing the short-time events also has the effect of removing small-amplitude fluctuations near the origin of the octants, so that the surviving sequences reflect large-amplitude energetic processes. All surviving three-node sub-sequences have similar average amplitudes.

The difference in temporal scales between the high and low $K_{OL}$ cases in the MFU2200 is minimal. The low $K_{OL}$ autocovariance curves for $r$ and $m$ are slightly closer to the FB180 curves, though this trend is not consistently observed across all three modes, and both high and low $K_{OL}$ MFU2200 curves are close to the FB180 curve for the $r$ and $s$ autocovariance and $T^+_{seq}$ p.d.f., so it is difficult to draw strong conclusions from these differences. While the motif and transition probability results show clear similarities between the low $K_{OL}$ near-wall dynamics and those of the FB180, this correspondence is not strongly reflected in the temporal scales. This may indicate that temporal scales are less sensitive to outer-layer conditioning than the structure of dynamical pathways: even when the outer layer is quiescent, its residual influence may be sufficient to maintain the faster timescales characteristic of high Reynolds number flow, while the specific sequences of dynamical transitions revert to patterns more typical of lower Reynolds number near-wall turbulence. Quantitatively determining the effect of outer-layer energy on near-wall temporal scales requires further analysis and is left for future work.

\section{Discussion}
\label{sec:discussion}


The central finding of this work is that the dynamical pathways sustaining near-wall turbulence are not fixed but depend on the flow configuration and, critically, on the energy state of the outer layer. State space and motif analysis reveal that while the SSP is visible in the mean trajectory across all three channels, it accounts for only a subset of the significant dynamical patterns. The exchanges of energy that drive turbulence follow multiple pathways, and the relative importance of these pathways shifts depending on what interactions are present.



Several non-SSP motifs play a significant role in turbulent dynamics across all three configurations. In particular, $(-r,+s,+m) \rightarrow (+r,+s,+m) \rightarrow (+r,-s,+m)$ is significant in all three channels. This motif describes an increase in roll-like mode energy leading to an overall high-energy state, followed by a decay in streak mode energy, consistent with a streak bursting event \citep{jimenez2005characterization, park2018bursting}. The universality of this motif across configurations with fundamentally different background flows suggests that streak bursting is a robust feature of near-wall turbulence that does not depend on the presence or absence of intra- or inter-layer interactions, but is instead an intrinsic property of the near-wall cycle itself. Despite streaks being the longest-lived of the three structures, their breakdown follows a regular, repeated pattern that is distinct from the background turbulence. The pattern $(+r,+s,-m) \rightarrow (+r,+s,+m) \rightarrow (-r,+s,+m)$, the most significant in the MFU180 and third most significant in the FB180, similarly involves the overall high-energy state but is preceded by an increase in meander mode energy and followed by a decrease in roll-like mode energy — a variation of a burst event where energetic meandering structures trigger roll destruction rather than streak breakdown.


The MFU180 and FB180 share common significant motifs, and their least significant motifs are all back-and-forth transitions. The under-representation of these back-and-forth patterns indicates that near-wall turbulence does not sustain itself through rapid oscillation between pairs of states, but instead progresses through longer sequences of distinct dynamical configurations. These longer sequences contain processes that do not correspond with the steps of the SSP, even in the dynamically restricted MFU180.


The present analysis does not identify any significant motif corresponding to the step of the SSP where rolls generate streaks, which may be attributed to a limitation of the POD modal basis rather than an absence of this process in the flow. The streak mode identified here contains non-negligible contributions from the wall-normal and spanwise velocity components, which are also the components most directly associated with roll-driven streak amplification. Similarly, the identified roll-like mode contains significant streamwise velocity fluctuations, which are associated with streaks. As a result, roll and streak dynamics are not cleanly separated in the POD basis, making it difficult to isolate the roll-to-streak transfer as a distinct motif. A decomposition that better separates these contributions, such as a resolvent-based or physically constrained modal basis, would be better suited to identifying this step of the SSP and is an important direction for future work.




The results on outer-layer conditioning represent the most important finding of this work. Prior work has established that large-scale outer structures modulate the amplitude of near-wall fluctuations \citep{hutchins2007evidence, mathis2009large}, but the present results show that their influence goes further: they fundamentally alter which dynamical pathways are active. When $K_{OL}$ is low, the near-wall dynamics of the MFU2200 closely resemble those of the FB180, sharing significant motifs and showing no tendency to oscillate between pairs of states. This is consistent with the findings of \cite{mathis2013estimating}, who showed that when the outer layer is quiescent its large-scale footprint on the near-wall region is reduced, effectively decoupling the inner and outer layer dynamics. The present results extend this picture by showing that the decoupling is not merely statistical but dynamical: the specific sequences of energy exchange among near-wall structures revert to those of a flow without a distinct outer layer. When $K_{OL}$ is high, the significant motifs are entirely distinct, with four of the five most significant containing the overall high-energy state $(+r,+s,+m)$, reflecting intermittent bursting events driven by outer-layer forcing. Since trajectories are computed by projecting the flow field onto POD modes without subtracting or filtering large-scale outer structures, we do not distinguish the effects of footprinting and amplitude modulation, and determining the relative importance of these mechanisms is left for future work. In contrast to low $K_{OL}$, the least significant motifs under high $K_{OL}$ are not dominated by back-and-forth transitions, suggesting that energetic outer-layer structures can occasionally drive the near-wall region back to a previous state — a disruption that does not occur when the outer layer is quiescent.

These findings have direct implications for turbulence modeling. Models based solely on SSP dynamics will fail to capture the alternate energy exchange pathways identified here, particularly those associated with bursting and outer-layer forcing. This is especially consequential for wall-modeled LES and other approaches that rely on near-wall dynamics to set boundary conditions for the outer flow, since the near-wall dynamical regime shifts substantially depending on the outer-layer energy state. Motif analysis provides a systematic framework for identifying and quantifying these pathways, and could inform the development of models that adapt their near-wall dynamics to the outer-layer state. In particular, the conditioning results suggest that a wall model that distinguishes between quiescent and energetic outer-layer states could better capture the correct near-wall dynamical regime in each case, potentially improving predictions of intermittent bursting and its contribution to skin friction. Extending this framework to higher Reynolds numbers would test whether the universality of the burst motif persists as the separation between inner and outer scales increases, and whether new dynamical pathways emerge that are not present at the Reynolds numbers studied here. Application to boundary layers, where the outer layer is less constrained than in channel flow, would further clarify the role of intermittent large-scale events in driving near-wall dynamics.


\section{Conclusions}
\label{sec:conclusion}





This work presents a view of turbulent dynamics as a finite library of dynamical pathways, and motif identification is used to identify the pathways that are critical for sustaining turbulence. Similar motifs emerge across Reynolds numbers, indicating that critical dynamical pathways exhibit a degree of universality. Conditioning analysis reveals that energetic outer-layer structures fundamentally modify the preferred dynamical pathways.

We have identified and quantified dynamical patterns using a state space trajectory framework combined with network motif analysis. DNS of three channel configurations, an MFU at $Re_\tau \approx 180$, a full-scale channel at $Re_\tau \approx 180$, and an MFU at $Re_\tau \approx 2200$, isolate the effects of intra-layer and inter-layer interactions on near-wall dynamics. POD modes corresponding to rolls, streaks, and meanders are used to define a three-dimensional state space, and the trajectory of the flow through this space is analyzed using motif identification to extract statistically overrepresented dynamical sequences.

The mean state space trajectory reproduces the SSP cycle across all three configurations, but instantaneous transition probabilities show that SSP-related transitions are not preferentially selected at any given moment. This reflects the SSP as a weak statistical drift rather than a dominant instantaneous pathway, and motivates the use of motif analysis to access the instantaneous dynamics directly.

Motif analysis reveals that while most sub-sequences occur with frequencies consistent with their pairwise transition statistics, a small number of patterns are significantly overrepresented or underrepresented, reflecting key dynamical processes. Some of these correspond to SSP steps, but several do not, indicating that dynamical processes beyond the SSP play an important role in sustaining near-wall turbulence even in the most idealized configuration. The MFU180 and FB180 share common significant motifs, confirming that the dominant near-wall dynamical processes are broadly similar across configurations with and without intra-layer interactions. The most universally significant motif, $(-r,+s,+m) \to (+r,+s,+m) \to (+r,-s,+m)$, appears in all three channels and is consistent with a streak bursting event, suggesting that this process is an intrinsic feature of near-wall turbulence that persists regardless of the flow configuration. The underrepresented motifs in the MFU180 and FB180 are back-and-forth transitions, indicating that near-wall turbulence does not sustain itself through rapid oscillation between pairs of states but instead progresses through longer sequences of distinct dynamical configurations.

Conditioning the MFU2200 results on outer-layer turbulent kinetic energy $K_{OL}$ reveals that the energy state of the outer layer fundamentally alters which dynamical pathways are active in the near-wall region. When $K_{OL}$ is low, the near-wall dynamics closely resemble those of the FB180, sharing significant motifs and exhibiting the same back-and-forth avoidance, consistent with a quiescent outer layer that decouples from the near-wall dynamics. When $K_{OL}$ is high, the significant motifs are entirely distinct, with four of the five most significant containing the overall high-energy state $(+r,+s,+m)$, reflecting intermittent bursting events driven by outer-layer forcing. These results show that the influence of the outer layer goes beyond amplitude modulation of near-wall fluctuations, fundamentally changing the sequences of energy exchange among near-wall structures.

Temporal scale analysis confirms that near-wall dynamics in the MFU180 are slower than in the FB180 and MFU2200, with structures persisting longer in the absence of intra-layer interactions. Both the slower dynamics and the elevated motif significance scores in the MFU180 reflect the same underlying constraint: the restricted domain size and low Reynolds number suppresses intra- and inter-layer interactions, reducing the variety of accessible dynamical pathways and causing certain recurrent patterns to appear more prominently than the FB180 and MFU2200. 

The state space trajectory and motif analysis framework developed here provides a systematic way to identify and quantify dynamical processes in turbulence beyond what time-averaged statistics alone reveal. The results have direct implications for turbulence modeling, particularly for wall-modeled LES, where capturing the correct near-wall dynamical regime under varying outer-layer states is essential for accurate prediction of bursting and skin friction fluctuations. Extensions of this framework to higher Reynolds numbers and other flow configurations such as boundary layers will further test the universality of the dynamical patterns identified here and clarify how they evolve with flow complexity.

\backsection[Acknowledgements]{The authors thank Prof. Adri{\'a}n Lozano-Dur{\'a}n for insightful discussions and suggestions, and  Prof. Javier Jim{\'e}nez for organizing the 2023 Madrid Turbulence Summer Workshop.}

\backsection[Funding]{This work was supported by the National Science Foundation Graduate Research Fellowship Program under grant no. 2139433 (E.L.), NASA's Transformational Tools and Technologies project under grant no. 80NSSC20M0201 (A.E.), and the European Research Council (ERC) under the Caust grant ERC-AdG101018287. Computations were performed on the Bridges-2 supercomputer at the Pittsburgh Supercomputing Center through allocation MCH250005 from the Advanced Cyberinfrastructure Coordination Ecosystem: Services \& Support (ACCESS) program, which is supported by U.S. National Science Foundation grants \#2138259; \#2138286; \#2138307; \#2137603; and \#2138296 \citep{boerner2023access}.}

\backsection[Declaration of interests]{The authors report no conflict of interest.}


\backsection[Author ORCIDs]{E. Lenz, https://orcid.org/0009-0001-0287-2967; A. Elnahhas, https://orcid.org/0000-0003-0449-9307; H.J. Bae, https://orcid.org/0000-0001-6789-6209}


\appendix
\section{Production-dissipation state space analysis}\label{appA}

Over the sub-domains, we track the turbulent production ($P$) and dissipation ($\varepsilon$), defined as
\begin{equation}
    \label{eq:proddiss}
    P=-\overline{uv\left(\frac{\partial U}{\partial y}\right)} \ , \ \varepsilon=\nu\left( \overline{-\mathbf{u}\cdot\boldsymbol{\nabla^2}\mathbf{u}}+{\nabla}^2K\right) \ .
\end{equation}

Production and dissipation form a trajectory through a two-dimensional state space. 

\begin{figure}
     \centering
     \begin{subfigure}[b]{0.4\textwidth}
         \centering
         \caption{}
         \vspace{-.2cm}
         \includegraphics[width=\textwidth,trim={3cm 0 3cm .9cm},clip]{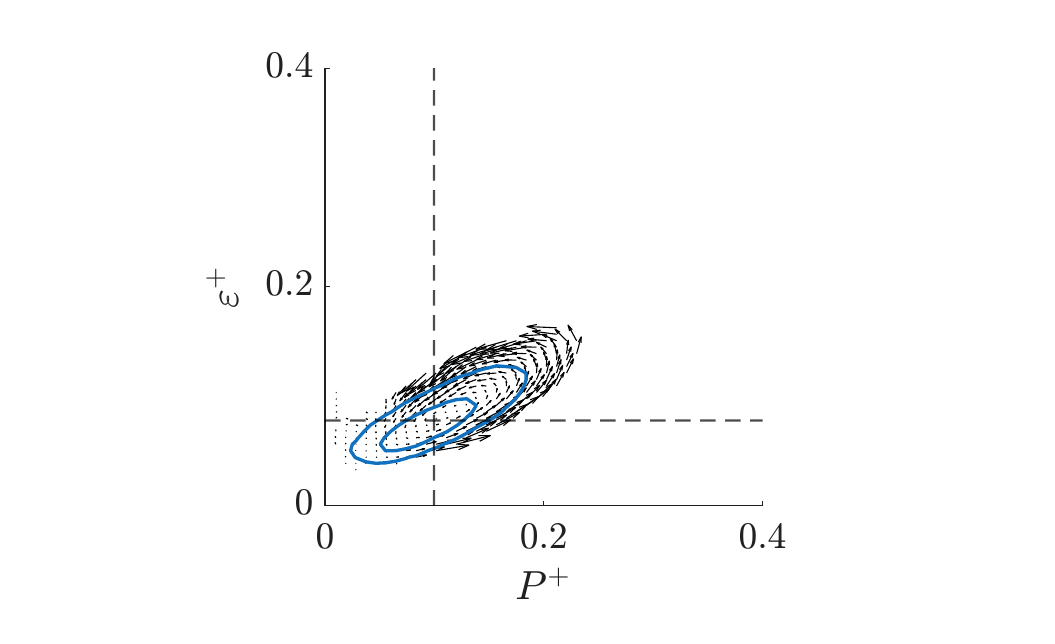}
     \end{subfigure}
     \begin{subfigure}[b]{0.4\textwidth}
         \centering
         \caption{}
         \vspace{-.2cm}
         \includegraphics[width=\textwidth,trim={3cm 0 3cm .9cm},clip]{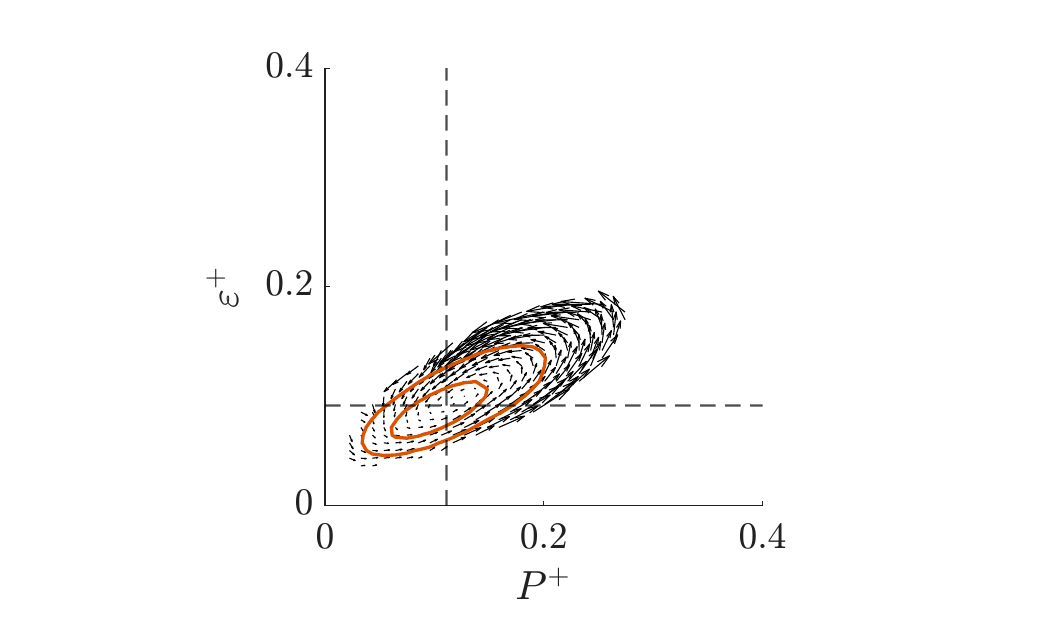}
     \end{subfigure}
        \caption{Production-dissipation state space for the (a) MFU180 and (b) FB180. Contours contain 90\% and 50\% of occurrences, and arrows show the average speed and direction of trajectory development.}
        \label{fig:mfu_pd_traj}
\end{figure}

Figure \ref{fig:mfu_pd_traj} shows the regions of state space that are occupied most frequently, reflecting common pairs of values of $P$ and $\varepsilon$, and the average trajectory through the $(P,\varepsilon)$ state space for the MFU180 and FB180. In the MFU180, a clear cycle is visible in the average trajectories, where increases (or decreases) in production precede increases (or decreases) in dissipation. The FB180 state space shows the same trend as the MFU180, though it spans a wider range of values of $P^+$ and $\varepsilon^+$ and the trajectory moves through the state space more quickly, indicated by the longer arrows in figure \ref{fig:mfu_pd_traj}(b). This suggests that the lack of periodicity and presence of intra-layer interactions in the FB180 do not fundamentally change the cycles of production and dissipation, but they increase the speed and intensity of the cycles. Data in the FB180 was collected over advecting sub-domains, so the fact that similar cycles are observed in the FB180 and MFU180 supports the use of the correlation-based advection technique for capturing dynamical cycles in the near-wall region.

We divide the state space into quadrants based on the median values of $P^+$ and $\varepsilon^+$, and we denote the quadrants with $+$ or $-$ when the quantity is above or below the median, consistent with the notation used for the $(r,s,m)$ octants. 
In quadrant $(-P,-\varepsilon)$ the mean trajectory develops slowly and spans a narrow range of values, indicating that when both production and dissipation are low, the channel undergoes slow, predictable dynamical processes. In contrast, in quadrant $(+P,+\varepsilon)$ the mean trajectory quickly moves through a wide range of production and dissipation values. In quadrant $(+P,-\varepsilon)$ the mean trajectory moves directly from a state of slow, low-amplitude dynamics to a state of fast, high-amplitude dynamics, and quadrant $(-P,+\varepsilon)$ does the opposite. Overall, the mean trajectory follows a pattern of quadrants: $(-P,-\varepsilon)\to(+P,-\varepsilon)\to(+P,+\varepsilon)\to(-P,+\varepsilon)\to(-P,-\varepsilon)$. This pattern captures a cycle where increases (or decreases) in production precede increases (or decreases) in dissipation, confirming that energy must be produced before it can be dissipated.



\begin{figure}
    \begin{flushleft}
        \hspace{1cm}
        (a)
        \hspace{3.9cm}
        (b)      
        \hspace{3.9cm}
        (c)
    \end{flushleft}
     \centering
     \begin{subfigure}[b]{0.32\textwidth}
         \centering
         \includegraphics[width=\textwidth,trim={2cm 0 3cm .9cm},clip]{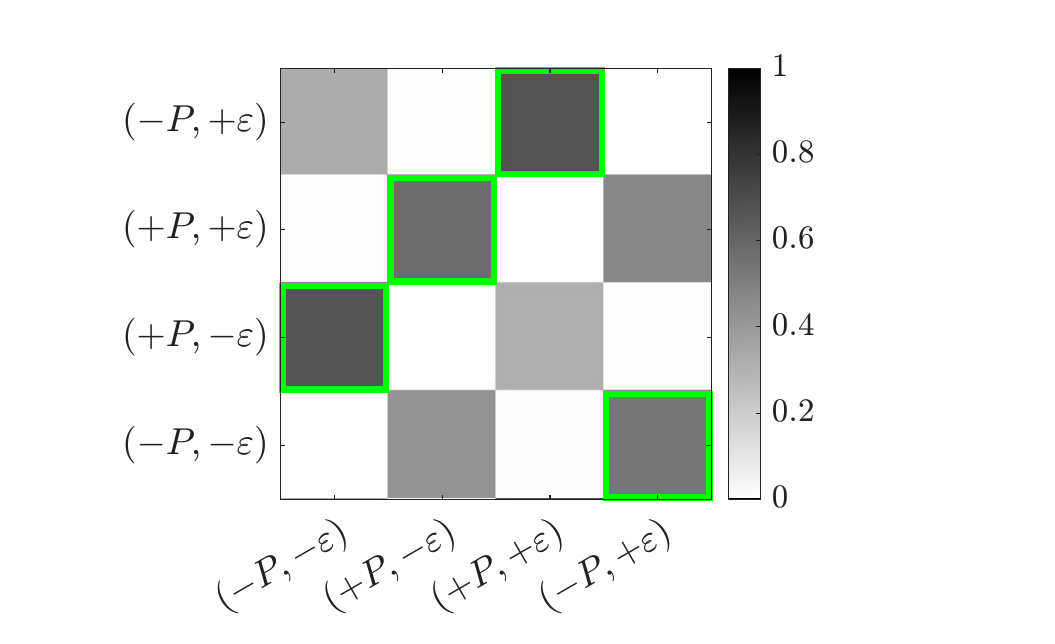}
     \end{subfigure}
     \begin{subfigure}[b]{0.32\textwidth}
         \centering
         \includegraphics[width=\textwidth,trim={2cm 0 3cm .9cm},clip]{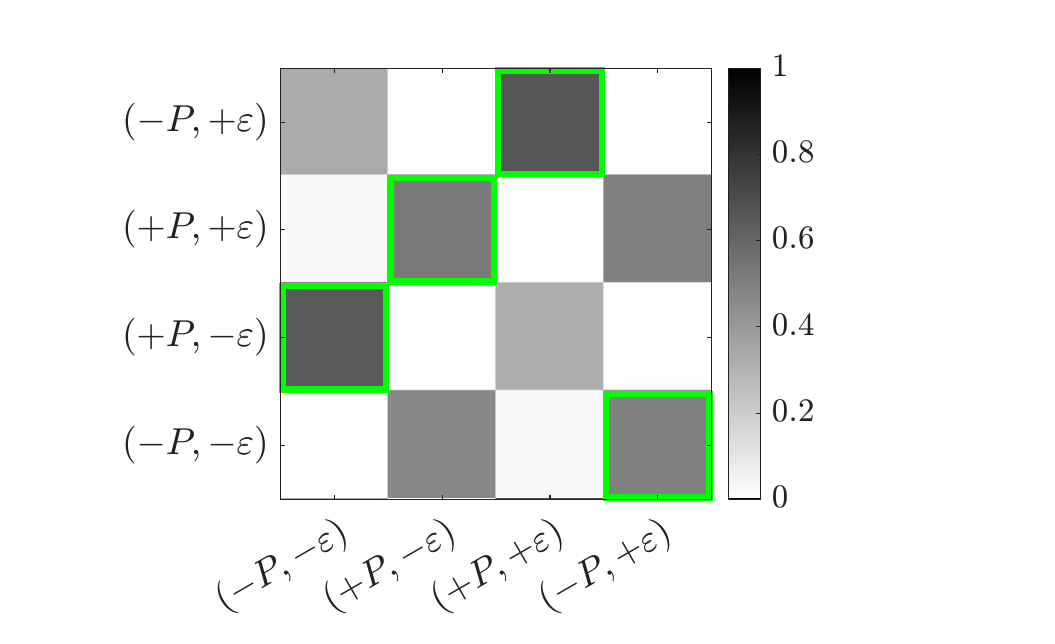}
     \end{subfigure}
     \begin{subfigure}[b]{0.32\textwidth}
         \centering
         \includegraphics[width=\textwidth,trim={2cm 0 3cm .9cm},clip]{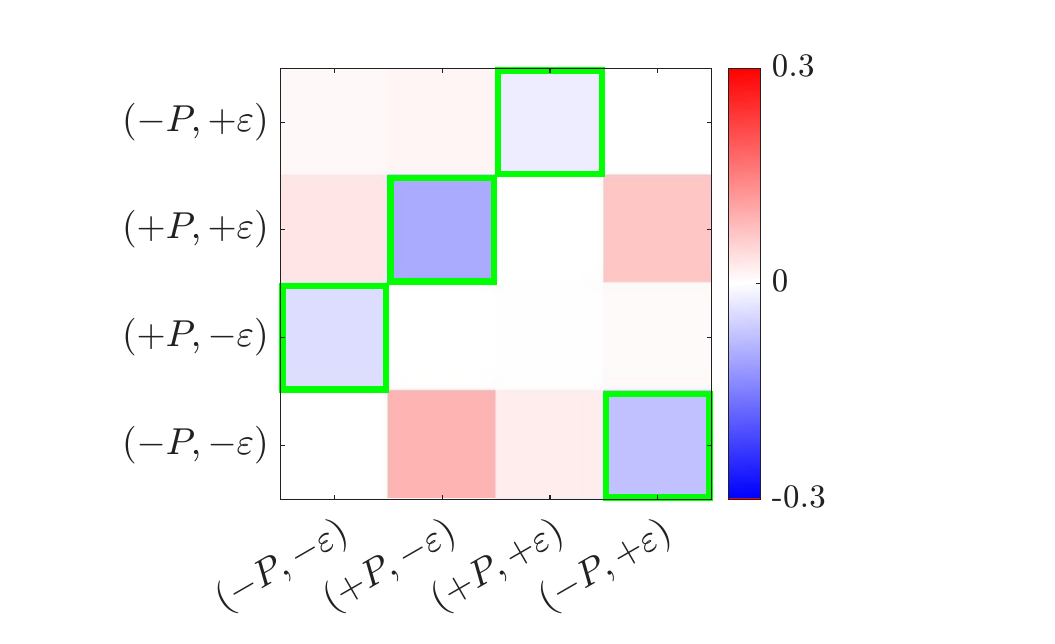}
     \end{subfigure}
        \caption{Production-dissipation transition probability for the (a) MFU180 and (b) FB180 from the horizontal axis to the vertical axis. (c) Difference between FB180 and MFU180 transition probabilities, where positive indicates a higher probability in the FB180. Green boxes highlight transitions forming the cycle of production preceding dissipation.}
        \label{fig:mfu_pd_trans}
\end{figure}

The transition probabilities between quadrants are calculated in the same way as the $(r,s,m)$ state-space and are visualized in figure \ref{fig:mfu_pd_trans}. The number of times the trajectories move between diagonally-adjacent quadrants (for example, $(-P,-\varepsilon)\to(+P,+\varepsilon)$) is negligibly small. We also observe that the transitions following the cycle of production preceding dissipation have high probabilities, capturing the same trend observed in the mean trajectory. However, transitions that do not follow the cycle of production preceding dissipation, i.e. $(+P,-\varepsilon)\to(-P,-\varepsilon)$, $(-P,-\varepsilon)\to(-P,+\varepsilon)$, $(-P,+\varepsilon)\to(+P,+\varepsilon)$, and $(+P,+\varepsilon)\to(+P,-\varepsilon)$, have notable transition probabilities. While production precedes dissipation for the majority of transitions, there are a significant number of instances where the opposite happens.

Figure \ref{fig:mfu_pd_trans}(b) shows the transition probabilities for the FB180, and overall they are similar to the MFU180. To visualize differences between the FB180 and MFU180, we subtract their transition probabilities, as shown in figure \ref{fig:mfu_pd_trans}(c). The FB180 has a slightly increased presence of transitions between diagonally-adjacent quadrants. Since the trajectory in the FB180 develops faster, it is able to pass through larger distances of the state-space between timesteps, creating diagonal transitions. Interestingly, the FB180 has a decreased number of transitions following the cycle of production preceding dissipation, and an increased number of transitions where dissipation precedes production. This may be because the sub-domains where $P^+$ and $\varepsilon^+$ are calculated in the FB180 do not span the whole channel, so energetic structures may drift in and out of the sub-domains. Further, it may suggest different dynamics between the channels. In the FB180, turbulence can be sustained by many interactions between various coherent structures, so the direct causal relationship between production and dissipation is less pronounced.

\begin{figure}
    \begin{flushleft}
        \hspace{1.3cm}
        (a)
        \hspace{3.6cm}
        (b)      
        \hspace{3.9cm}
        (c)
    \end{flushleft}
     \centering
     \begin{subfigure}[b]{0.32\textwidth}
         \centering
         \includegraphics[width=\textwidth,trim={2cm 0 3cm .9cm},clip]{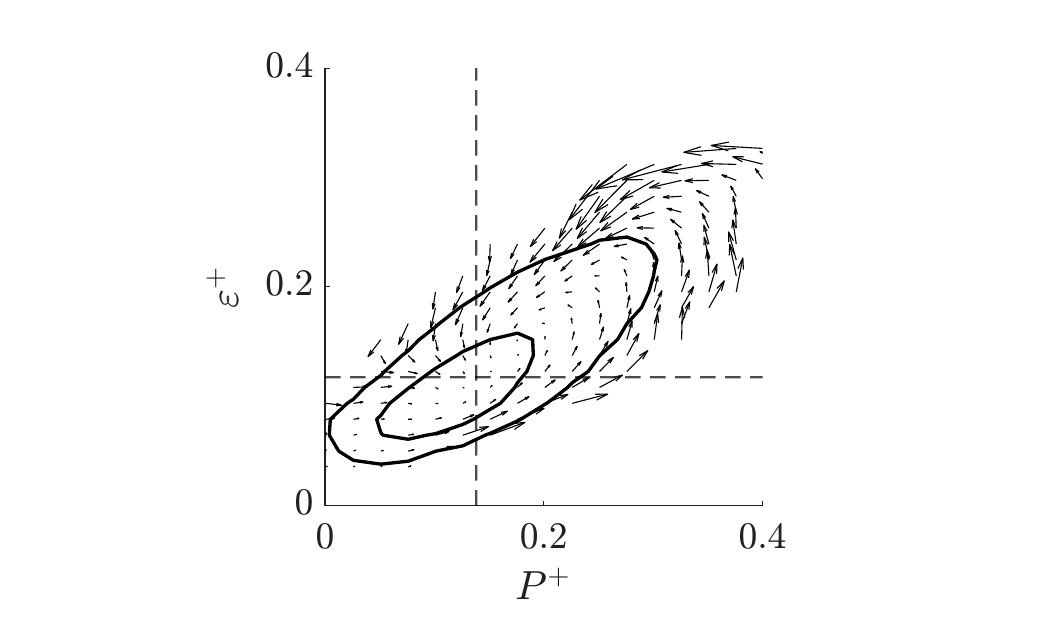}
     \end{subfigure}
     \begin{subfigure}[b]{0.32\textwidth}
         \centering
         \includegraphics[width=\textwidth,trim={2cm 0 3cm .9cm},clip]{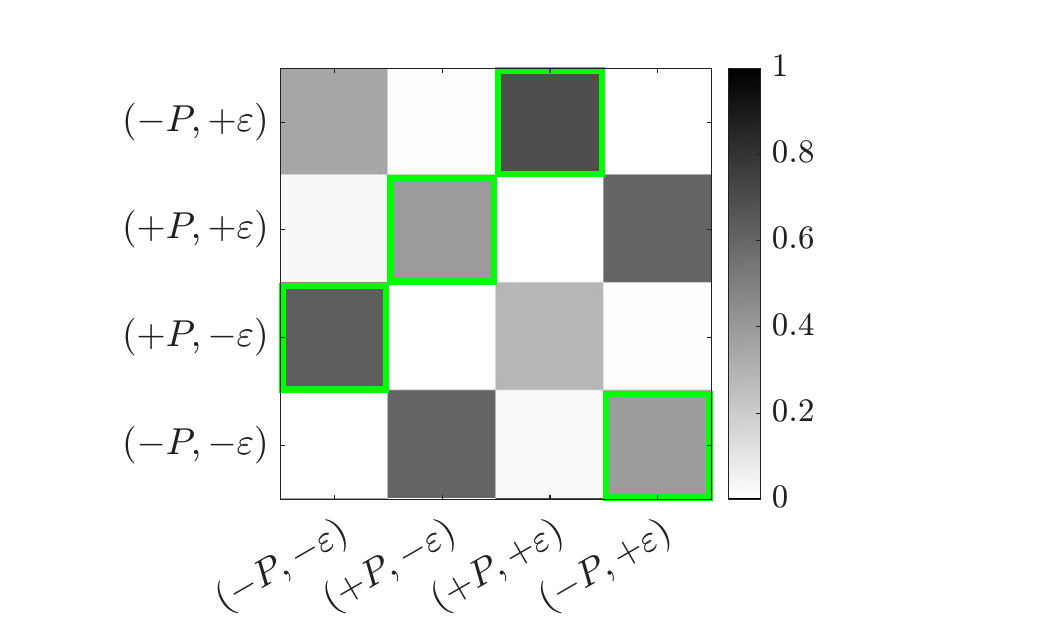}
     \end{subfigure}
     \begin{subfigure}[b]{0.32\textwidth}
         \centering
         \includegraphics[width=\textwidth,trim={2cm 0 3cm .9cm},clip]{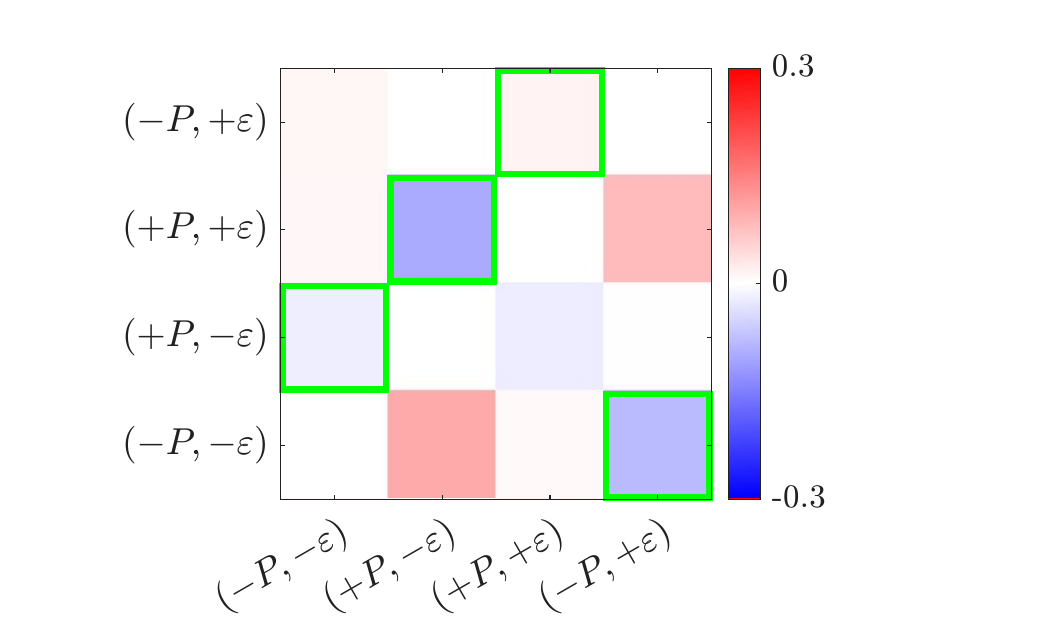}
     \end{subfigure}
        \caption{(a) Production-dissipation state space for the MFU2200. Contours contain 90\% and 50\% of occurrences, and arrows show the average speed and direction of trajectory development. (b) Production-dissipation transition probability for the MFU2200. Shade of the square represents the probability of moving from the octant on the horizontal axis to the octant on the vertical axis. (c) Difference between MFU2200 and MFU180 transition probabilities, where positive indicates a higher probability in the MFU2200. Green boxes highlight transitions forming the cycle of production preceding dissipation.}
        \label{fig:llmfu_pd_traj}
\end{figure}

Figure \ref{fig:llmfu_pd_traj} shows the production-dissipation state space for the full MFU2200 dataset, which is affected by intra-layer interactions between near-wall structures, and inter-layer interactions between near-wall structures and outer-layer structures. As shown in \ref{fig:llmfu_pd_traj}(a), a cycle of production preceding dissipation is visible in the mean trajectory, and the trajectory spans a wider range of the state space than in the MFU180 or FB180. Similar to the effect of intra-layer interactions, inter-layer interactions further increase the speed and amplitude of the mean trajectory, though the fundamental cycle is still present.

Figure \ref{fig:llmfu_pd_traj}(b) shows the transition probability, and in contrast to the MFU180 and FB180, the cycle of production preceding dissipation is not pronounced. Instead, transitions involving increases or decreases in $P^+$ have higher probability than transitions involving changes in $\varepsilon^+$. This lack of a cycle can also be seen in the discrepancy between the MFU2200 and MFU180 transition probabilities, where transitions related to the cycle of production preceding dissipation appear less in the MFU2200 than the MFU180. This observation indicates that large structures facilitate the transfer of turbulent kinetic energy between different spatial regions. Further, the large structures contain regions of high and low production and dissipation, which may extend into the near-wall region and interfere with cycles of production and dissipation \cite{lee2019spectral}. 
The discrepancy between the MFU2200 and MFU180 is shown in figure \ref{fig:llmfu_pd_traj}(c), and it is similar to the FB180's discrepancy in figure \ref{fig:mfu_pd_trans}, though with a larger magnitude, suggesting that the larger structures present in the MFU2200 are capable of transferring more energy spatially and significantly disrupting the cycle of production preceding dissipation.

\section{Derivation of $\alpha$}\label{appB}

To compute the nonlinear energy exchange between POD modes, we follow the procedure of \cite{ding2025mode}, but while this study computed energy exchange between Fourier modes at a specified wall-normal location, we extend the derivation to capture energy exchanged between POD modes, which vary in $y$. 

 We begin by finding the turbulent kinetic equation for energy contained in each mode. In this derivation, we abbreviate general velocity fields $\mathbf{u}(x,y,z,t)$ as $\mathbf{u}$ and general Fourier-transformed velocity fields $\widehat{\mathbf{u}}(k_x,k_z;y,t)$ as $\widehat{\mathbf{u}}$. Recalling that the subscript $q$ is an abbreviation for a function of $(k_{x,q},k_{z,q},n_q)$, the energy equation for a POD mode $q$ is derived by

 \begin{equation}
     \frac{\partial E_q}{\partial t} = \frac{\partial \left( \frac{1}{2}\beta_q(t)\beta^*_q(t)\right)}{\partial t} = \mathrm{Real}\left\{\beta^*_q(t) \frac{\partial \beta_q(t)}{\partial t} \right\} = \mathrm{Real}\left\{ \left\langle \frac{\partial \left(\beta_q(t) \widehat{\boldsymbol{\phi}}_{q}(y)\right)}{\partial t} \ , \ \beta_q(t) \widehat{\boldsymbol{\phi}}_{q}(y) \right\rangle \right\} \ .
 \end{equation}


By orthogonality, the term $\frac{\partial \left(\beta_q(t) \widehat{\boldsymbol{\phi}}_{q}(y)\right)}{\partial t}$ can be replaced with the derivative of the velocity field $\frac{\partial \widehat{\mathbf{u}}}{\partial t}$, which can be expressed with the momentum equation: 

\begin{equation}
    \label{eq:modeenergy2}
    \frac{\partial E_{q}}{\partial t} = \mathrm{Real}\left\{\left\langle \frac{\partial \widehat{\mathbf{u}}}{\partial t} \ , \ \beta_q(t)\widehat{\boldsymbol{\phi}}_{q}(y)\right\rangle \ \right\} = \mathrm{Real}\left\{\left\langle \mathscr{L}\left( \widehat{\mathbf{u}} \right) - \ \mathcal{F} \left\{ (\widehat{\mathbf{u}} \cdot \nabla) \widehat{\mathbf{u}} \right\} \ , \ \beta_q(t)\widehat{\boldsymbol{\phi}}_{q}(y) \right\rangle \right\} .
\end{equation}

Here, $\mathcal{F}\{(\cdot)\}$ denotes a Fourier transform in $x$ and $z$ and is equivalent to $\widehat{(\cdot)}$. $\mathscr{L}$ captures the pressure gradient and viscosity diffusion terms of the momentum equation. When the inner product is performed on $\mathscr{L}\left( \widehat{\mathbf{u}} \right)$, the only nonzero contributions will be from structures with wavenumbers $(k_x=k_{x,q} \ , \ k_z=k_{z,q})$ by orthogonality.

In this work we focus on the nonlinear term, since it represents energy exchange between structures of different scales. Only contributions from terms with $(k_x=k_{x,q} \ , \ k_z=k_{z,q})$ result in a nonzero inner product. Since the nonlinear advection term $(\widehat{\mathbf{u}} \cdot \nabla) \widehat{\mathbf{u}}$ is a product, its resulting wavenumber is a convolution of the wavenumbers of its terms, which are denoted $(k_{x,l},k_{z,l})$ and $(k_{x,q}-k_{x,l},k_{z,q}-k_{z,l})$:


\begin{equation}
    \mathcal{F} \left\{ (\widehat{\mathbf{u}} \cdot \nabla) \widehat{\mathbf{u}}  \right\}(k_{x,q},k_{z,q}) = \sum\limits_{\substack{k_{x,\ell},k_{z,\ell}}} \left(\widehat{\mathbf{u}}(k_{x,q}-k_{x,\ell}, k_{z,q}-k_{z,\ell}) \cdot \nabla  \right) \widehat{\mathbf{u}}(k_{x,\ell},k_{z,\ell}) \ .
\end{equation}

By expressing the flow fields as sums of POD modes and their temporally-varying coefficients, i.e. $\widehat{\mathbf{u}}(k_x, k_z; y,t) = \sum\limits_{n=1}^N \beta(k_x,k_z,n;t) \widehat{\boldsymbol{\phi}}(k_x,k_z,n;y)$, we express the nonlinear energy exchange between different POD modes. With this substitution, and isolating the nonlinear term of equation \ref{eq:modeenergy2}, we have

\begin{equation}
    \label{eq:nonlinearenergyeq}
    \begin{split}
    \frac{\partial E_{q, \ nonlin.}}{\partial t} &= \mathrm{Real}\left\{ \left\langle -\mathcal{F} \left\{ (\widehat{\mathbf{u}} \cdot \nabla) \widehat{\mathbf{u}} \right\} \ , \ \beta_q(t)\widehat{\boldsymbol{\phi}}_{q}  \right\rangle \right\} \\
    &= - \mathrm{Real}\left\{\left\langle    \sum\limits_{\substack{k_{x,\ell},k_{z,\ell}}} \sum\limits_{n_l=1}^N \sum\limits_{n_p=1}^N \left( \beta_p(t)\widehat{\boldsymbol{\phi}}_p(y) \cdot \nabla \right)\left(\beta_\ell(t) \widehat{\boldsymbol{\phi}}_\ell(y) \right) \ , \ \beta_q(t)\widehat{\boldsymbol{\phi}}_{q}(y) \ \right\rangle \right\} ,
    \end{split}
\end{equation}

where the subscript $\ell$ is abbreviation for quantities that are a function of $(k_{x,\ell}, k_{z,\ell},n_\ell)$, and the subscript $p$ is abbreviation for quantities that are a function of $(k_{x,q}-k_{x,\ell}, k_{z,q}-k_{z,\ell},n_p)$.

We rearrange equation \ref{eq:nonlinearenergyeq} to isolate the contributions from specific modes $\ell$ and $p$ to mode $q$, which we denote ${\alpha}(\ell,p,q;t)$:

\begin{equation}
\begin{split}
    \frac{\partial E_{q, \ nonlin.}}{\partial t} &=\sum\limits_{\substack{k_{x,l},k_{z,l}}} \sum\limits_{n_l=1}^N \sum\limits_{n_p=1}^N - \mathrm{Real}\left\{\beta_\ell(t) \beta_p(t) \beta_q^*(t) \left\langle \left( \widehat{\boldsymbol{\phi}}_p(y) \cdot \nabla \right) \widehat{\boldsymbol{\phi}}_\ell(y) \ , \ \widehat{\boldsymbol{\phi}}_{q}(y)  \right\rangle \right\} \\ &=  \sum\limits_{\substack{k_{x,\ell},k_{z,\ell}}} \sum\limits_{n_l=1}^N \sum\limits_{n_p=1}^N {\alpha}(\ell,p,q;t) \ ,
\end{split}
\end{equation}

where $\alpha(\ell,p,q;t)$ is defined as

\begin{equation}
     \alpha(\ell,p,q;t)  = -\mathrm{Real} \left\{ \beta_\ell(t) \beta_p(t) \beta_q^*(t) \left\langle \left( \widehat{\boldsymbol{\phi}}_p(y) \cdot \nabla \right) \widehat{\boldsymbol{\phi}}_\ell(y) \ , \ \widehat{\boldsymbol{\phi}}_{q}(y) \right\rangle \right\} \ .
\end{equation}

In this formulation, ${\alpha}(\ell,p,q;t)$ reflects the energy transferred from mode $\ell$ to mode $q$, with mode $p$ facilitating the interaction. By rearranging the entries, ${\alpha}(p,\ell,q;t)$ would reflect the energy moving from mode $p$ to mode $q$, with mode $\ell$ facilitating the interaction. By conservation of energy, ${\alpha}(\ell,p,q;t)$ = -${\alpha}(q,p^*,\ell;t)+\textit{flux}$, where the symmetry is broken due to energy flux crossing the top or sides of the domain.  



\section{Sensitivity study of significance scores}\label{appC}


To ensure that the identified motifs genuinely capture underlying dynamics, we investigate parameters that may affect the significance scores $\widetilde{W}(\mathbf{s})$ of the sub-sequences. We observe the effect of the number of randomly generated sequences, the cutoff frequency of the reported sub-sequences, and the removal of short-time events.

Since motif significance score calculation involves collecting statistical quantities over randomly generated sequences ($\overline{f_R(\mathbf{s}})$ and $Var(f_R(\mathbf{s}))$), we perform a sensitivity study to ensure that these statistics are converged. Intermediate significance scores are calculated using half of the randomly generated sequences. At this intermediate stage, the ranked order of the sub-sequences is the same as the final result, verifying that generating additional random sequences would not substantially change the results. The average discrepancy between the intermediate and final significance scores is 0.84\%.

\begin{figure}
     \centering
     \begin{subfigure}[b]{0.49\textwidth}
         \centering
         \caption{}
         \includegraphics[width=\textwidth,trim={0 0 0 .8cm},clip]{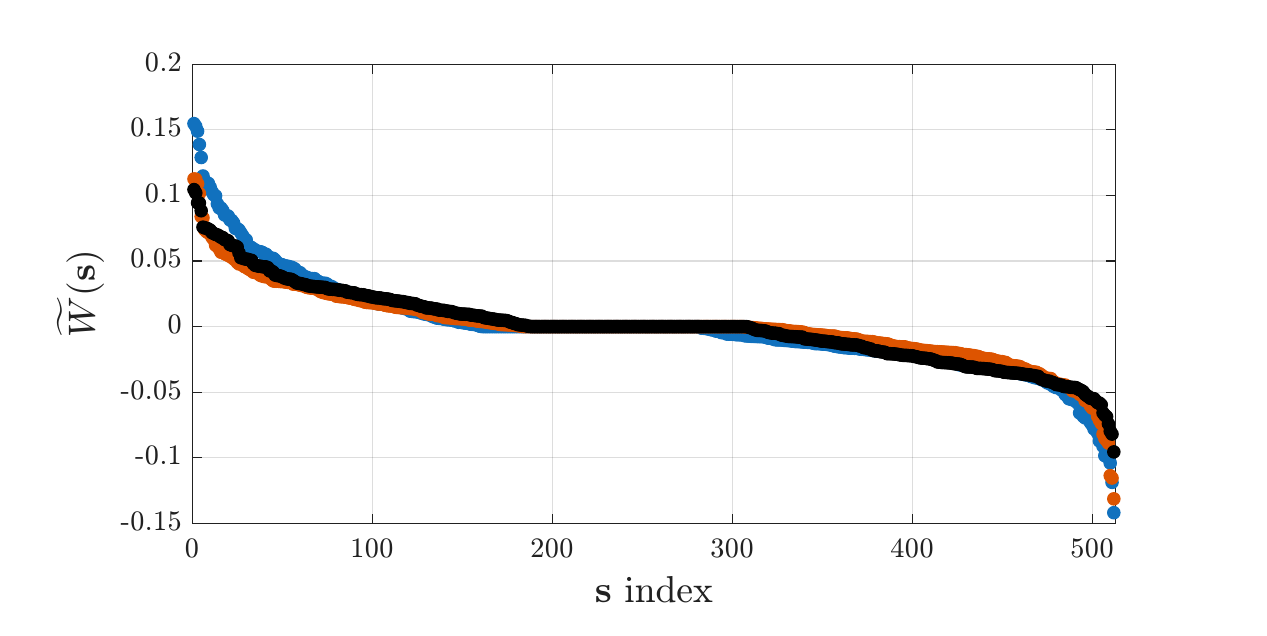}
     \end{subfigure}
     \begin{subfigure}[b]{0.49\textwidth}
         \centering
         \caption{}
         \includegraphics[width=\textwidth,trim={0 0 0 .8cm},clip]{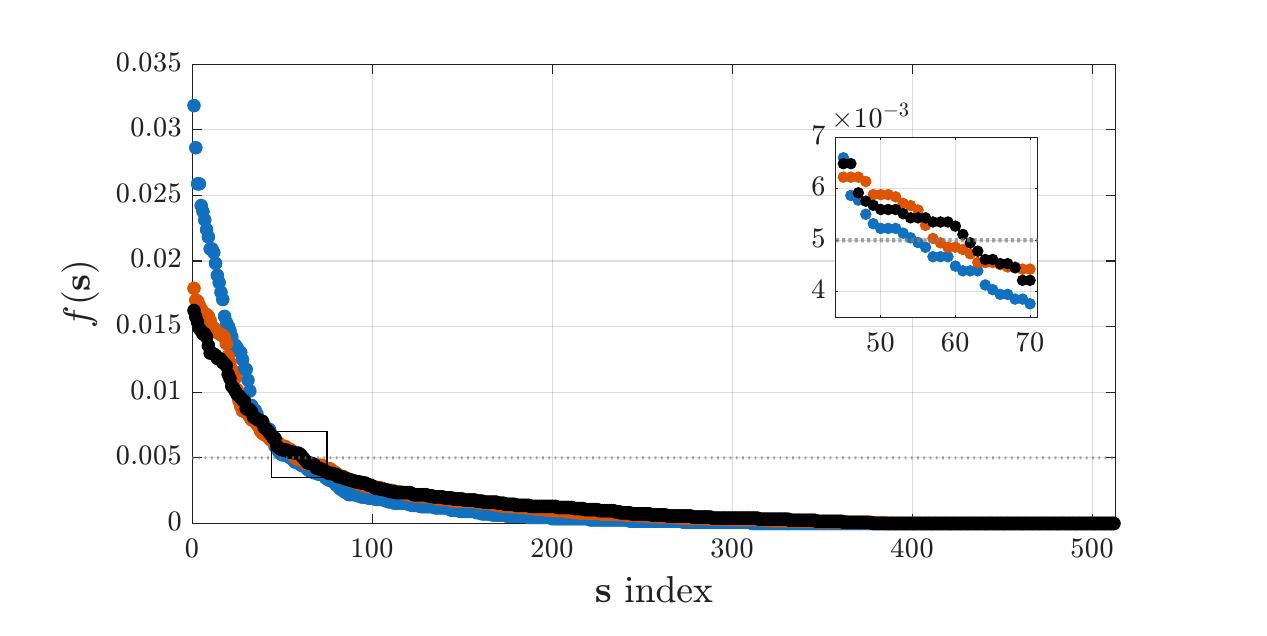}
     \end{subfigure}
        \caption{Significance scores of all sub-sequences of length three, sorted highest to lowest, for the MFU180 (blue), FB180 (red), and MFU2200 (black). (b) Frequencies of all sub-sequences of length three, sorted highest to lowest, for the same datasets. Grey dotted line indicates the cutoff value of $f(\mathbf{s}) = 0.005$. We note that the ranking order in (a) is different from the ranking order in (b).}
        \label{fig:sensitivitystudy}
\end{figure}

An artifact of motif significance calculation is that rare events produce large-magnitude significance scores. Rare sub-sequences manifest very infrequently in the randomly generated sequences, with many randomly generated sequences having zero instances of the rare sub-sequence. This results in a very low variance $Var(f_R(\mathbf{s}))$ and a high significance score. Therefore, we do not report the significance of sub-sequences that have frequencies below a cutoff value. Figure \ref{fig:sensitivitystudy}(a) shows the ranked significance scores of all 512 sub-sequences without removing low-frequency sub-sequences. Figure \ref{fig:sensitivitystudy}(b) shows the distribution of frequencies of all subsequences, and an inflection point is observed near $f(\mathbf{s})=0.005$, so we do not report significance scores of sub-sequences with $f(\mathbf{s})\leq0.005$.

To evaluate the effect of removing sequence events with $T_{seq}^+\leq5.3$, we repeat the significance score calculation on the trajectories without removing short-time events. Rare events (for example, crossing between diagonally-adjacent octants) are largely driven by short-time fluctuations, so their presence is increased in the unfiltered datasets. This exacerbates the issue of rare events producing high significance scores, especially in the FB180 and MFU2200. With the unfiltered FB180 and MFU2200 data, the most significant motifs are rare events, and the significance scores of the majority of the non-rare sub-sequences are negative. This does not reflect the underlying physical processes we wish to observe, so we report the results of the trajectories with short-time events removed.

\bibliographystyle{jfm}
\bibliography{jfm}

\end{document}